\documentclass[11pt, a4paper]{article}

\usepackage{ifthen}

\newcommand{\MyUniPat}{lsdfgkhjvrkjlhmisdlcjn}

\makeatletter

\newcommand{\NewcommandThreeArgsTwoOpt}[5]{
\DeclareRobustCommand#1{\@ifnextchar[%
{\csname\expandafter\@gobble\string#1@presq\endcsname}%
{\csname\expandafter\@gobble\string#1@nopresq\endcsname}}
\expandafter\def\csname\expandafter\@gobble\string#1@nopresq\endcsname##1{\@ifnextchar[%
{\csname\expandafter\@gobble\string#1@nopresq@postsq\endcsname[]{##1}}%
{\csname\expandafter\@gobble\string#1@nopresq@nopostsq\endcsname[]{##1}}}
\expandafter\def\csname\expandafter\@gobble\string#1@presq\endcsname[##1]##2{\@ifnextchar[%
{\csname\expandafter\@gobble\string#1@presq@postsq\endcsname[{##1}]{##2}}%
{\csname\expandafter\@gobble\string#1@presq@nopostsq\endcsname[{##1}]{##2}}}
\expandafter\def\csname\expandafter\@gobble\string#1@nopresq@nopostsq\endcsname[##1]##2{#2}
\expandafter\def\csname\expandafter\@gobble\string#1@presq@nopostsq\endcsname[##1]##2{#3}
\expandafter\def\csname\expandafter\@gobble\string#1@nopresq@postsq\endcsname[##1]##2[##3]{#4}
\expandafter\def\csname\expandafter\@gobble\string#1@presq@postsq\endcsname[##1]##2[##3]{#5}
}

\newcommand{\NewcommandFourArgsTwoOpt}[5]{
\DeclareRobustCommand#1{\@ifnextchar[%
{\csname\expandafter\@gobble\string#1@presq\endcsname}%
{\csname\expandafter\@gobble\string#1@nopresq\endcsname}}
\expandafter\def\csname\expandafter\@gobble\string#1@nopresq\endcsname##1##2{\@ifnextchar[%
{\csname\expandafter\@gobble\string#1@nopresq@postsq\endcsname[]{##1}{##2}}%
{\csname\expandafter\@gobble\string#1@nopresq@nopostsq\endcsname[]{##1}{##2}}}
\expandafter\def\csname\expandafter\@gobble\string#1@presq\endcsname[##1]##2##3{\@ifnextchar[%
{\csname\expandafter\@gobble\string#1@presq@postsq\endcsname[{##1}]{##2}{##3}}%
{\csname\expandafter\@gobble\string#1@presq@nopostsq\endcsname[{##1}]{##2}{##3}}}
\expandafter\def\csname\expandafter\@gobble\string#1@nopresq@nopostsq\endcsname[##1]##2##3{#2}
\expandafter\def\csname\expandafter\@gobble\string#1@presq@nopostsq\endcsname[##1]##2##3{#3}
\expandafter\def\csname\expandafter\@gobble\string#1@nopresq@postsq\endcsname[##1]##2##3[##4]{#4}
\expandafter\def\csname\expandafter\@gobble\string#1@presq@postsq\endcsname[##1]##2##3[##4]{#5}
}

\NewcommandFourArgsTwoOpt{\ifndef}
{\@ifundefined{#2}{#3}{}}
{\ClassError{My Macros}{The first optional argument is not expected in "\ifndef"}{}}
{\@ifundefined{#2}{#3}{#4}}
{\ClassError{My Macros}{The first optional argument is not expected in "\ifndef"}{}}

\NewcommandThreeArgsTwoOpt{\myusepackage}
{\@ifpackageloaded{#2}{}{\usepackage{#2}}}
{\@ifpackageloaded{#2}{}{\usepackage[#1]{#2}}}
{\@ifpackageloaded{#2}{}{#3 \usepackage{#2}}}
{\@ifpackageloaded{#2}{}{#3 \usepackage[#1]{#2}}}

\makeatother

\myusepackage{amssymb}
\myusepackage{amsmath}  %

\myusepackage{amsthm}[ ]  %

\myusepackage[T1,OT2,T2A]{fontenc} 
\DeclareTextSymbolDefault{\CYRYAT}{OT2}
\DeclareTextSymbolDefault{\cyryat}{OT2}
\DeclareTextSymbolDefault{\CYRFITA}{OT2}
\DeclareTextSymbolDefault{\cyrfita}{OT2}
\DeclareTextSymbolDefault{\CYRIZH}{OT2}
\DeclareTextSymbolDefault{\cyrizh}{OT2}
\myusepackage[utf8]{inputenc}

\usepackage[greek,ukrainian,russian,english]{babel}

\myusepackage{soul}      %
\setul{1.25pt}{.16pt}  %
\myusepackage{latexsym}
\myusepackage{amsfonts}
\myusepackage{units}    %
\newcommand{\dr}{\nicefrac}
\myusepackage[nolabel]{fnbreak}  %
\myusepackage{psfrag}   %
\myusepackage[hyphens,spaces,obeyspaces]{url}   %
 
\myusepackage{hyperref}  %
\setpdflinkmargin{0.75pt}  %

\myusepackage[usenames,dvipsnames]{xcolor}
\hypersetup{
colorlinks,
linkcolor={red!55!black},
citecolor={blue!55!black},
urlcolor={blue!35!black}
}

\myusepackage{hyphenat}  %
\myusepackage{caption}   %
\myusepackage{microtype}   %
\myusepackage{marginnote}  %
\myusepackage{calc}  %
\myusepackage{mathrsfs}   %
\myusepackage{enumitem}  %
\myusepackage{graphicx}

\newcommand{\dgCapDefinition}{Definition}
\newcommand{\dgCapDefinitions}{Definitions}
\newcommand{\dgCapPostulate}{Postulate}
\newcommand{\dgCapPostulates}{Postulates}
\newcommand{\dgCapExample}{Example}

\newcommand{\dgCapFact}{Fact}
\newcommand{\dgCapFacts}{Facts}
\newcommand{\dgCapQuestion}{Question}
\newcommand{\dgCapQuestions}{Questions}
\newcommand{\dgCapLemma}{Lemma}
\newcommand{\dgCapLemmas}{Lemmas}
\newcommand{\dgCapNotation}{Notation}
\newcommand{\dgCapCorollary}{Corollary}
\newcommand{\dgCapCorollaries}{Corollaries}
\newcommand{\dgCapProposition}{Proposition}
\newcommand{\dgCapPropositions}{Propositions}
\newcommand{\dgCapClaim}{Claim}
\newcommand{\dgCapClaims}{Claims}
\newcommand{\dgCapTheorem}{Theorem}
\newcommand{\dgCapTheorems}{Theorems}
\newcommand{\dgCapProblem}{Problem}
\newcommand{\dgCapProblems}{Problems}
\newcommand{\dgCapRemark}{Remark}
\newcommand{\dgCapRemarks}{Remarks}
\newcommand{\dgCapConjecture}{Conjecture}
\newcommand{\dgCapConjectures}{Conjectures}
\newcommand{\dgCapResult}{Result}

\newcommand{\dgCapPart}{Part}
\newcommand{\dgCapParts}{Parts}
\newcommand{\dgCapChapter}{Chapter}
\newcommand{\dgCapChapters}{Chapters}
\newcommand{\dgCapSection}{Section}
\newcommand{\dgCapSections}{Sections}
\newcommand{\dgCapSubsection}{Subsection}
\newcommand{\dgCapSubsections}{Subsections}
\newcommand{\dgCapFigure}{Figure}
\newcommand{\dgCapFigures}{Figures}
\newcommand{\dgCapEquation}{Equation}
\newcommand{\dgCapEquations}{Equations}
\newcommand{\dgCapExpression}{Expression}
\newcommand{\dgCapExpressions}{Expressions}
\newcommand{\dgCapInequality}{Inequality}
\newcommand{\dgCapInequalities}{Inequalities}
\newcommand{\dgProofOf}{\proofname\ of}

\newcommand{\dgDefinition}{Definition}
\newcommand{\dgDefinitions}{Definitions}
\newcommand{\dgPostulate}{Postulate}
\newcommand{\dgPostulates}{Postulates}

\newcommand{\dgFact}{Fact}
\newcommand{\dgFacts}{Facts}
\newcommand{\dgQuestion}{Question}
\newcommand{\dgQuestions}{Questions}
\newcommand{\dgLemma}{Lemma}
\newcommand{\dgLemmas}{Lemmas}
\newcommand{\dgCorollary}{Corollary}
\newcommand{\dgCorollaries}{Corollaries}
\newcommand{\dgProposition}{Proposition}
\newcommand{\dgPropositions}{Propositions}
\newcommand{\dgClaim}{Claim}
\newcommand{\dgClaims}{Claims}
\newcommand{\dgTheorem}{Theorem}
\newcommand{\dgTheorems}{Theorems}
\newcommand{\dgProblem}{Problem}
\newcommand{\dgProblems}{Problems}
\newcommand{\dgRemark}{Remark}
\newcommand{\dgRemarks}{Remarks}
\newcommand{\dgConjecture}{Conjecture}
\newcommand{\dgConjectures}{Conjectures}

\newcommand{\dgPart}{Part}
\newcommand{\dgParts}{Parts}
\newcommand{\dgChapter}{Chapter}
\newcommand{\dgChapters}{Chapters}
\newcommand{\dgSection}{Section}
\newcommand{\dgSections}{Sections}
\newcommand{\dgSubsection}{Subsection}
\newcommand{\dgSubsections}{Subsections}
\newcommand{\dgFigure}{Figure}
\newcommand{\dgFigures}{Figures}
\newcommand{\dgEquation}{Equation}
\newcommand{\dgEquations}{Equations}
\newcommand{\dgExpression}{Expression}
\newcommand{\dgExpressions}{Expressions}
\newcommand{\dgInequality}{Inequality}
\newcommand{\dgInequalities}{Inequalities}

\ifndef{theorem}{}
\ifndef{theorem*}{\newtheorem*{theorem*}{\dgCapTheorem}}

\ifndef{lemma}{}
\ifndef{lemma*}{\newtheorem*{lemma*}{\dgCapLemma}}
\ifndef{corollary}{}
\ifndef{corollary*}{\newtheorem*{corollary*}{\dgCapCorollary}}
\ifndef{conjecture}{}
\ifndef{conjecture*}{\newtheorem*{conjecture*}{\dgCapConjecture}}
\ifndef{proposition}{}
\ifndef{proposition*}{\newtheorem*{proposition*}{\dgCapProposition}}
\ifndef{claim}{}
\ifndef{claim*}{\newtheorem*{claim*}{\dgCapClaim}}
\ifndef{result}{}
\ifndef{result*}{\newtheorem*{result*}{\dgCapResult}}
\ifndef{problem}{}
\ifndef{problem*}{\newtheorem*{problem*}{\dgCapProblem}}

\newtheoremstyle{mydefinition}  %
{\topsep}{\topsep}  %
{\slshape}  %
{}  %
{\bfseries}  %
{.}  %
{ }  %
{}  %

\newtheoremstyle{mynotation}  %
{\topsep}{\topsep}  %
{}  %
{}  %
{\bfseries\itshape}  %
{.}  %
{ }  %
{}  %

\newtheoremstyle{myremark}  %
{\topsep}{\topsep}  %
{\slshape}  %
{}  %
{\bfseries\itshape}  %
{.}  %
{ }  %
{\thmname{#1}\thmnumber{~#2}}  %

\newtheoremstyle{myexample}  %
{\topsep}{\topsep}  %
{\itshape}  %
{}  %
{\slshape}  %
{.}  %
{ }  %
{\underline{\thmname{#1}\thmnumber{~#2}}}  %

\newtheoremstyle{myclaims}  %
{\topsep}{\topsep}  %
{\slshape}  %
{}  %
{\bfseries\slshape}  %
{.}  %
{ }  %
{\thmname{#1}\thmnumber{~#2}\thmnote{\textnormal{~(#3)}}}  %

\ifndef{notation}{\theoremstyle{mynotation}}

{\theoremstyle{myremark}

\newtheorem*{myremark*}{\dgCapRemark}
}
{\theoremstyle{mydefinition}
\ifndef{postulate}{}
\ifndef{postulate*}{\newtheorem*{postulate*}{\dgCapPostulate}}
\ifndef{definition}{}
\ifndef{definition*}{\newtheorem*{definition*}{\dgCapDefinition}}
}
{\theoremstyle{myexample}
\ifndef{example}{}
\ifndef{example*}{\newtheorem*{example*}{\dgCapExample}}
}
{\theoremstyle{myclaims}

\newtheorem*{my_claim*}{\dgCapClaim}
\ifndef{fact}{}
\ifndef{fact*}{\newtheorem*{fact*}{\dgCapFact}}
\ifndef{question}{}
\ifndef{question*}{\newtheorem*{question*}{\dgCapQuestion}}
}

\newtheoremstyle{anystatementst}  %
{\topsep}{\topsep}  %
{\itshape}  %
{}  %
{\bfseries}  %
{.}  %
{ }  %
{#3}  %

{\theoremstyle{anystatementst} }

\newcommand{\newident}[3][\MyUniPat]{\ifthenelse{\equal{\MyUniPat}{#1}}
{
\newcommand{#2}[1][]{\ensuremath{\mathit{#3##1}}}
}
{\ifthenelse{\equal{}{#1}}
{
\newcommand{#2}[1][]{\ensuremath{\mathit{#3}}}
}
{
\newcommand{#2}[1][\MyUniPat]{\ifthenelse{\equal{\MyUniPat}{##1}}%
{\ensuremath{\mathit{#1}}}%
{\ensuremath{\mathit{#3}}}}
}
}
}

\newcommand{\newidenT}[3][\MyUniPat]{\ifthenelse{\equal{\MyUniPat}{#1}}
{
\newcommand{#2}[1][\MyUniPat]{\ifthenelse{\equal{\MyUniPat}{##1}}%
{\il{#3}}%
{\ensuremath{\mathit{#3##1}}}}
}
{
\newcommand{#2}[1][\MyUniPat]{\ifthenelse{\equal{\MyUniPat}{##1}}%
{\il{#1}}%
{\ensuremath{\mathit{#3}}}}
}
}

\newcommand{\newmat}[3][\MyUniPat]{\ifthenelse{\equal{\MyUniPat}{#1}}%
{\newcommand{#2}[1][]{#3##1}}%
{\newcommand{#2}[1][]{#3}}%
}

\newcommand{\providemat}[3][\MyUniPat]{\ifthenelse{\equal{\MyUniPat}{#1}}
{\providecommand{#2}[1][]{#3##1}}
{\providecommand{#2}[1][]{#3}}  %
}

\newcommand{\newmatop}[3][\MyUniPat]{\ifthenelse{\equal{\MyUniPat}{#1}}
{
\newcommand{#2}{\operatorname{#3}}
}
{
\newcommand{#2}[1][\MyUniPat]{\ifthenelse{\equal{\MyUniPat}{##1}}%
{\operatorname{#1}}%
{\operatorname{#3}}}
}
}

\newcommand{\newmatoparg}[3][\MyUniPat]{\ifthenelse{\equal{\MyUniPat}{#1}}
{
\newcommand{#2}[1]{\ifthenelse{\equal{}{##1}}{\operatorname{#3}}{\operatorname{#3}\l(##1\r)}}
}
{
\newcommand{#2}[2][\MyUniPat]{\ifthenelse{\equal{\MyUniPat}{##1}}%
{\ifthenelse{\equal{}{##2}}{\operatorname{#1}}{\operatorname{#1}\l(##2\r)}}%
{\ifthenelse{\equal{}{##2}}{\operatorname{#3}}{\operatorname{#3}\l(##2\r)}}}
}
}

\newcommand{\newOlike}[2]{
\newcommand{#1}[2][\MyUniPat]{\ifthenelse{\equal{\MyUniPat}{##1}}%
{\ensuremath{\mathit{#2}\lf(##2\rt)}}%
{#2(##2)}%
}}

\makeatletter
\newcommand{\MyMakeTheoMacros}[3]{
\expandafter\newcommand\csname\expandafter\@gobble\string#2NostarNoname@DGaux\endcsname[2][]
{\ifthenelse{\equal{}{##1}}%
{\begin{#1}~##2 \end{#1}}%
{\begin{#1}\label{##1}~##2\end{#1}}%
}
\expandafter\newcommand\csname\expandafter\@gobble\string#2StarNoname@DGaux\endcsname[1]
{\begin{#1*}~##1 \end{#1*}}
\newcommand#2{\expandafter\@ifstar%
\expandafter{\csname\expandafter\@gobble\string#2StarNoname@DGaux\endcsname}%
{\csname\expandafter\@gobble\string#2NostarNoname@DGaux\endcsname}%
}

\expandafter\newcommand\csname\expandafter\@gobble\string#2NostarName@DGaux\endcsname[3][]
{\ifthenelse{\equal{}{##1}}%
{\begin{#1}[\e{##2}]~##3 \end{#1}}%
{\begin{#1}[\e{##2}]\label{##1}~##3\end{#1}}%
}
\expandafter\newcommand\csname\expandafter\@gobble\string#2StarName@DGaux\endcsname[2]
{\begin{#1*}[\e{##1}]~##2 \end{#1*}}
\newcommand#3{\expandafter\@ifstar%
\expandafter{\csname\expandafter\@gobble\string#2StarName@DGaux\endcsname}
{\csname\expandafter\@gobble\string#2NostarName@DGaux\endcsname}%
}
}

\newtheorem*{rep@theorem}{\rep@title}
\newcommand{\newreptheorem}[2]{%
\newenvironment{rep#1}[1]{%
\def\rep@title{#2 \ref{##1}}%
\begin{rep@theorem}}%
{\end{rep@theorem}}}
\makeatother

\newcommand{\MyMakeDupTheoMacros}[7]{
\MyMakeTheoMacros{#1}{#2}{#3}
\newreptheorem{#1}{#6}
\newcommand{#4}[3]{
\newcommand{##2}{##3}
\begin{#1}\label{##1}~##2\end{#1}}
\newcommand{#5}[4]{
\newcommand{##2}{##4}
\begin{#1}{\e{##3}}\label{##1}~##2\end{#1}}
\newcommand{#7}[2]{\begin{rep#1}{##1}~##2 \end{rep#1}}
}

\newcommand{\MyMakeRefMacros}[3]{\newcommand{#1}[2][]
{\ifthenelse{\equal{}{##1}}{#2~\ref{##2}}{#3~\ref{##1} and~\ref{##2}}}}

\newcommand{\MyMakeEqRefMacros}[3]{\newcommand{#1}[2][]
{\ifthenelse{\equal{}{##1}}{#2~\eqref{##2}}{#3~\eqref{##1} and~\eqref{##2}}}}

\MyMakeTheoMacros{fact}{\fct}{\nfct}

\MyMakeRefMacros{\fctref}{\dgFact}{\dgFacts}
\MyMakeRefMacros{\Fctref}{\dgCapFact}{\dgCapFacts}

\MyMakeTheoMacros{question}{\quest}{\nquest}

\MyMakeRefMacros{\questref}{\dgQuestion}{\dgQuestions}
\MyMakeRefMacros{\Questref}{\dgCapQuestion}{\dgCapQuestions}

\MyMakeTheoMacros{notation}{\nota}{\nnota}

\MyMakeDupTheoMacros{lemma}
{\lem}{\nlem}{\lemdup}{\nlemdup}{\dgLemma}{\lemrep}

\MyMakeRefMacros{\lemref}{\dgLemma}{\dgLemmas}
\MyMakeRefMacros{\Lemref}{\dgCapLemma}{\dgCapLemmas}

\MyMakeDupTheoMacros{corollary}
{\crl}{\ncrl}{\crldup}{\ncrldup}{\dgCorollary}{\crlrep}

\MyMakeRefMacros{\crlref}{\dgCorollary}{\dgCorollaries}
\MyMakeRefMacros{\Crlref}{\dgCapCorollary}{\dgCapCorollaries}

\MyMakeTheoMacros{proposition}{\prp}{\nprp}

\newtheorem*{prp*}{\e{\dgCapProposition}}

\MyMakeRefMacros{\prpref}{\dgProposition}{\dgPropositions}
\MyMakeRefMacros{\Prpref}{\dgCapProposition}{\dgCapPropositions}

\MyMakeDupTheoMacros{my_claim}
{\clm}{\nclm}{\clmdup}{\nclmdup}{\dgClaim}{\clmrep}

\MyMakeRefMacros{\clmref}{\dgClaim}{\dgClaims}
\MyMakeRefMacros{\Clmref}{\dgCapClaim}{\dgCapClaims}

\MyMakeDupTheoMacros{theorem}
{\theo}{\ntheo}{\theodup}{\ntheodup}{\dgTheorem}{\theorep}

\MyMakeRefMacros{\theoref}{\dgTheorem}{\dgTheorems}
\MyMakeRefMacros{\Theoref}{\dgCapTheorem}{\dgCapTheorems}

\MyMakeTheoMacros{postulate}{\postu}{\npostu}

\MyMakeRefMacros{\posturef}{\dgPostulate}{\dgPostulates}
\MyMakeRefMacros{\Posturef}{\dgCapPostulate}{\dgCapPostulates}

\MyMakeTheoMacros{definition}{\defi}{\ndefi}

\MyMakeRefMacros{\defiref}{\dgDefinition}{\dgDefinitions}
\MyMakeRefMacros{\Defiref}{\dgCapDefinition}{\dgCapDefinitions}

\MyMakeTheoMacros{problem}{\prob}{\nprob}

\MyMakeRefMacros{\probref}{\dgProblem}{\dgProblems}
\MyMakeRefMacros{\Probref}{\dgCapProblem}{\dgCapProblems}

\MyMakeTheoMacros{myremark}{\rem}{\nrem}

\MyMakeRefMacros{\remref}{\dgRemark}{\dgRemarks}
\MyMakeRefMacros{\Remref}{\dgCapRemark}{\dgCapRemarks}

\MyMakeTheoMacros{conjecture}{\conj}{\nconj}

\MyMakeRefMacros{\conjref}{\dgConjecture}{\dgConjectures}
\MyMakeRefMacros{\Conjref}{\dgCapConjecture}{\dgCapConjectures}

\renewcommand{\qedsymbol}{$\blacksquare$}

\newcommand{\prfstart}[1][]{\ifthenelse{\equal{}{#1}}%
{\begin{proof}\renewcommand{\qedsymbol}{$\blacksquare$}}%
{\begin{proof}[\dgProofOf\ #1]%
\renewcommand{\qedsymbol}{$\blacksquare_{\mbox{\it{\scriptsize{#1}}}}$}}%
}
\newcommand{\prfend}[1][*]{%
\ifthenelse{\equal{}{#1}}{\renewcommand{\qedsymbol}{$\blacksquare$}}{}%
\ifthenelse{\equal{*}{#1}}{}%
{\renewcommand{\qedsymbol}{$\blacksquare_{\mbox{\it{\scriptsize{#1}}}}$}}%
\end{proof}\renewcommand{\qedsymbol}{$\blacksquare$}%
}

\newcommand{\NewSectLikeDG}[3]{
\NewcommandThreeArgsTwoOpt{#1}
{\ifthenelse{\equal{*}{##2}}{#2*}{#2{##2}}}
{#2{##2\label{##1}}}
{#2[##3]{##2#3{##3}}#3{##3}}
{#2[##3]{##2#3{##3}\label{##1}}#3{##3}}
}

\NewSectLikeDG{\sect}{\section}{\sectionmark}

\NewSectLikeDG{\ssect}{\subsection}{\subsectionmark}

\NewSectLikeDG{\sssect}{\subsubsection}{\subsubsectionmark}

\makeatletter
\@addtoreset{chapter}{part}
\makeatother

\newcommand*\parttitle{}
\let\origpart\part
\renewcommand*{\part}[2][]{%
\ifx\\#1\\
\origpart{#2}%
\renewcommand*\parttitle{#2}%
\else
\origpart[#1]{#2}%
\renewcommand*\parttitle{#1}%
\fi
}

\ifndef{chapter}
{\NewSectLikeDG{\chap}{\part}{\DoNothing}}  %
[\NewSectLikeDG{\chap}{\chapter}{\chaptermark}]

\NewSectLikeDG{\prt}{\part}{\DoNothing}  %

\MyMakeRefMacros{\prtref}{\dgPart}{\dgParts}
\MyMakeRefMacros{\Prtref}{\dgCapPart}{\dgCapParts}

\MyMakeRefMacros{\chref}{\dgChapter}{\dgChapters}
\MyMakeRefMacros{\Chref}{\dgCapChapter}{\dgCapChapters}

\MyMakeRefMacros{\sref}{\dgSection}{\dgSections}
\MyMakeRefMacros{\Sref}{\dgCapSection}{\dgCapSections}

\MyMakeRefMacros{\ssref}{\dgSubsection}{\dgSubsections}
\MyMakeRefMacros{\Ssref}{\dgCapSubsection}{\dgCapSubsections}

\MyMakeRefMacros{\sssref}{\dgSubsection}{\dgSubsections}
\MyMakeRefMacros{\Sssref}{\dgCapSubsection}{\dgCapSubsections}

\MyMakeRefMacros{\figref}{\dgFigure}{\dgFigures}
\MyMakeRefMacros{\Figref}{\dgCapFigure}{\dgCapFigures}

\newcommand{\IfMathMode}[2]{\ifmmode{#1}\else{#2}\fi}

\newcommand{\fbr}[1]{\IfMathMode%
{#1}{$#1$}}                     %

\newcommand{\fnbr}[1]{\mbox{\fbr{#1}}}  %

\newcommand{\fla}[2][*]{\ifthenelse{\equal{}{#1}}{\fbr{#2}}{\fnbr{#2}}}

\newcommand{\mal}[2][]{\MyChangeMathMargins
\delimiterfactor=1001 %
\ifthenelse{\equal{}{#1}}%
{\begin{align*} #2 \end{align*}}%
{\ifthenelse{\equal{P}{#1}}%
{\allowdisplaybreaks\begin{align*} #2%
\end{align*}\interdisplaylinepenalty=10000}%
{\begin{align}\begin{split}\label{#1} #2 \end{split}\end{align}}%
}\delimiterfactor=901%
}

\MyMakeEqRefMacros{\equref}{\dgEquation}{\dgEquations}
\MyMakeEqRefMacros{\Equref}{\dgCapEquation}{\dgCapEquations}

\MyMakeEqRefMacros{\expref}{\dgExpression}{\dgExpressions}
\MyMakeEqRefMacros{\Expref}{\dgCapExpression}{\dgCapExpressions}

\MyMakeEqRefMacros{\inequref}{\dgInequality}{\dgInequalities}
\MyMakeEqRefMacros{\Inequref}{\dgCapInequality}{\dgCapInequalities}

\makeatletter

\DeclareRobustCommand\bref{\@ifnextchar[%
{\bref@presq}%
{\bref@nopresq}%
}

\def\bref@presq[#1]{\@ifnextchar[%
{(\ref{#1}), \bref@presq}
{(\ref{#1}) and~\bref@nopresq}
}

\def\bref@nopresq#1{(\ref{#1})}

\newcommand\Cases{%
\left\{\!\!\!\begin{array}{ll}\Cases@continue%
}

\def\Cases@continue#1#2{\@ifnextchar\bgroup%
{#1 &\txt{#2}\\ \Cases@continue}%
{#1 &\txt{#2}\end{array}\Cases@end}%
}

\def\Cases@end{\@ifnextchar[%
{\Cases@end@postsq}%
{\right.}
}

\def\Cases@end@postsq[#1]{\ifthenelse{\equal{\}}{#1}}
{\!\!\right\}}%
{\right.}
}

\makeatother

\newcommand{\lf}{\mathopen{}\mathclose\bgroup\left}
\newcommand{\rt}{\aftergroup\egroup\right}

\providecommand{\middle}{\big}
\newcommand{\md}{\middle}

\newcommand{\ud}{\vphantom{|_1^1}}

\newcommand{\chs}{\genfrac(){0cm}{}}  %

\newmatop{\plog}{poly-log}
\newmatop[disc]{\disc}{disc_{#1}}
\newmatop{\Pow}{Pow}

\newmatoparg{\supp}{supp}   %
\newmatoparg{\diam}{diam}   %

\newcommand{\NewHLikeDG}[2]{
\NewcommandThreeArgsTwoOpt{#1}
{\operatorname{\mathnormal{#2}}\l(##2\r)}
{\operatorname{\mathnormal{#2_{##1}}}\l(##2\r)}
{\operatorname{\mathnormal{#2}}\l(##2\md|{\ud}##3\r)}
{\operatorname{\mathnormal{#2_{##1}}}\l(##2\md|{\ud}##3\r)}
}

\NewHLikeDG{\h}{H}

\NewHLikeDG{\hm}{H_{min}}

\newcommand{\NewILikeDG}[2]{
\NewcommandFourArgsTwoOpt{#1}
{\mathop{\pmb{#2}}\lf[##2:{\ud}##3\rt]}
{\mathop{\pmb{#2}\?\!_{##1}}\lf[##2:{\ud}##3\rt]}
{\mathop{\pmb{#2}}\lf[##2:##3\md|{\ud}##4\rt]}
{\mathop{\pmb{#2}\?\!_{##1}}\lf[##2:##3\md|{\ud}##4\rt]}
}

\NewILikeDG{\I}{I}

\newcommand{\NewELikeDG}[2]{
\NewcommandThreeArgsTwoOpt{#1}
{#2\lf[##2\rt]}
{#2_{##1}\lf[##2\rt]}
{#2\lf[##2\md|{\ud}##3\rt]}
{#2_{##1}\lf[##2\md|{\ud}##3\rt]}
}

\NewELikeDG{\PR}{\mathop{\pmb{Pr}}}

\NewELikeDG{\E}{\mathop{\pmb{E}}}

\NewELikeDG{\Del}{\mathop{\pmb{\Delta}}}

\NewELikeDG{\Var}{\mathop{\pmb{Var}}}

\providemat{\NN}{\mathbb{N}}

\newcommand{\wtl}{\widetilde}
\newcommand{\overbar}[1]{\mkern 1.5mu\overline{\mkern-1.5mu#1\mkern-1.5mu}\mkern 1.5mu}
\newcommand{\wbr}{\overbar}

\newcommand{\fr}[3][*]{%
\ifthenelse{\equal{*}{#1}}%
{\frac{#2}{#3}}{}%
\ifthenelse{\equal{}{#1}}%
{\dr{#2}{#3}}{}%
\ifthenelse{\equal{/}{#1}}%
{\lf.#2\md/#3\rt.}{}%
\ifthenelse{\equal{p_}{#1}}%
{\lf.\lf(#2\rt)\md/#3\rt.}{}%
\ifthenelse{\equal{_p}{#1}}%
{\lf.#2\md/\lf(#3\rt)\rt.}{}%
\ifthenelse{\equal{pp}{#1}}%
{\lf.\lf(#2\rt)\md/\lf(#3\rt)\rt.}{}%
}

\NewcommandThreeArgsTwoOpt{\set}
{\lf\{#2\rt\}}
{}
{\lf\{#2\md|{\ud}#3\rt\}}
{\lf\{#2{\ud}#1{\ud}#3\rt\}}

\newcommand{\s}{\set}

\newcommand{\Log}[2][]{\ifthenelse{\equal{}{#1}}%
{\log\lf(#2\rt)}%
{\log_{#1}\lf(#2\rt)}%
}

\newcommand{\NewMinLikeDG}[2]{
\NewcommandThreeArgsTwoOpt{#1}
{#2\lf\{##2\rt\}}
{#2_{##1}\lf\{##2\rt\}}
{#2\lf\{##2\md|\ud##3\rt\}}
{#2_{##1}\lf\{##2\md|\ud##3\rt\}}
}

\NewMinLikeDG{\Min}{\min}

\NewMinLikeDG{\Max}{\max}

\newmatop{\argmin}{argmin}
\NewMinLikeDG{\Argmin}{\argmin}

\newmatop{\argmax}{argmax}
\NewMinLikeDG{\Argmax}{\argmax}

\NewMinLikeDG{\Sup}{\sup}

\NewMinLikeDG{\Inf}{\inf}

\newOlike{\asO}{O}
\newOlike{\astO}{\wtl O}
\newOlike{\aso}{o}
\newOlike{\asOm}{\Omega}
\newOlike{\astOm}{\wtl\Omega}
\newOlike{\asom}{\omega}
\newOlike{\asT}{\Theta}
\newOlike{\astT}{\wtl\Theta}

\makeatletter

\DeclareRobustCommand\bra{\@ifnextchar[%
{\bra@presq}%
{\bra@nopresq}%
}

\def\bra@presq[#1]{\@ifnextchar[%
{\lla #1\rt|\bra@presq}
{\lla #1\rt|\bra@nopresq}
}

\def\bra@nopresq#1{\lla #1\rt|}

\DeclareRobustCommand\ket{\@ifnextchar[%
{\ket@presq}%
{\ket@nopresq}%
}

\def\ket@presq[#1]{\@ifnextchar[%
{\lf|#1\rra\ket@presq}
{\lf|#1\rra\ket@nopresq}
}

\def\ket@nopresq#1{\lf|#1\rra}

\makeatother

\newcommand{\sz}[2][]{\ifthenelse{\equal{}{#1}}%
{\lf|#2\rt|}%
{\lf|#2\rt|_{#1}}}

\providecommand{\ceil}[2][*]{\ifthenelse{\equal{}{#1}}%
{\lceil #2 \rceil}%
{\llc #2 \rrc}}

\newcommand{\txt}[1]{\textrm{#1}}  %

\DeclareMathAlphabet{\mathbfcal}{OMS}{cmsy}{b}{n}

\DeclareMathAlphabet{\mathlowcal}{OT1}{pzc}{m}{it}

\newidenT{\Pp}{P}
\newidenT[BPP]{\BPP}{BPP_{#1}}
\newidenT{\SBP}{SBP}
\newidenT{\PP}{PP}
\newidenT{\UPP}{UPP}
\newidenT{\NP}{NP}
\newidenT{\AM}{AM}
\newidenT[MA]{\MA}{MA_{#1}}
\newidenT[UAM]{\UAM}{UAM_{#1}}

\newidenT{\Disj}{Disj}
\newidenT{\IP}{IP}

\newmat{\mset}{\smin\set}

\newcommand{\lla}{\lf\langle}
\newcommand{\rra}{\rt\rangle}
\newcommand{\llc}{\lf\lceil}
\newcommand{\rrc}{\rt\rceil}

\newcommand{\To}{\Rightarrow}
\newcommand{\Then}{\Longrightarrow}

\newcommand{\dt}{\cdot}
\newcommand{\tm}{\cdot}
\newcommand{\deq}{\stackrel{\textrm{def}}{=}}
\newcommand{\smin}{\setminus}

\newcommand{\sbseq}{\subseteq}
\newcommand{\nsbseq}{\nsubseteq}
\newcommand{\sbs}{\subset}

\newcommand{\eps}{\varepsilon}

\newcommand{\overlay}[3][0mu]{
\begingroup
\mathchoice{
\bgroup
\ooalign{$\displaystyle#2$\cr
\hidewidth{$\displaystyle\mkern#1#3$}\hidewidth}
\egroup
}{
\bgroup
\ooalign{$\textstyle#2$\cr
\hidewidth{$\textstyle\mkern#1#3$}\hidewidth}
\egroup
}{
\bgroup
\ooalign{$\scriptstyle#2$\cr
\hidewidth{$\scriptstyle\mkern#1#3$}\hidewidth}
\egroup
}{
\bgroup
\ooalign{$\scriptscriptstyle#2$\cr
\hidewidth{$\scriptscriptstyle\mkern#1#3$}\hidewidth}
\egroup
}
\endgroup
}

\newcommand{\unin}{\mathrel{\overlay[1.5mu]{\subset}{\sim}}}

\makeatletter
\def\mov@rlay#1#2{\leavevmode\vtop{%
\baselineskip\z@skip \lineskiplimit-\maxdimen
\ialign{\hfil$\m@th#1##$\hfil\cr#2\crcr}}}
\makeatother

\newcommand{\ds}[1][]
{\ifthenelse{\equal{}{#1}}{\allowbreak\dots}{#1\allowbreak\dots#1}}
\newmat{\dc}{\ds[,]}

\newmat{\OI}{\s{0,1}} 

\mathchardef\myhyphen="2D

\newcommand{\g}[1]{%
\ifthenelse{\equal{#1}-}{\myhyphen}{}%
\ifthenelse{\equal{#1}=}{\equiv}{}%
\ifthenelse{\equal{#1}+}{\oplus}{}%
}

\makeatletter

\let\dgampersand\&

\DeclareRobustCommand\&{%
\new@ifnextchar[%
{\dgsep@reposit}%
{\dgampersand}%
}

\def\dgsep@reposit[#1]{\hspace{#1}&\hspace{-#1}}

\makeatother

\newcommand{\abstart}{\begin{abstract}}
\newcommand{\abend}{\end{abstract}}

\newenvironment{myepig}
{\par\addtolength{\leftskip}{28mm}\addtolength{\rightskip}{8mm}\noindent\ignorespaces}
{\par}

\newenvironment{myepigsgn}
{\par\addtolength{\leftskip}{84mm}\noindent\ignorespaces}
{\par}

\NewcommandThreeArgsTwoOpt{\epig}
{{\small \begin{myepig} \textit{#2} \end{myepig} \addvspace{\baselineskip}}}
{\ClassError{My Macros}{The first optional argument is not expected in "\epig"}{}}
{{\small \begin{myepig} \textit{#2} \end{myepig} \Nopagebreak%
\begin{myepigsgn} -- #3 \end{myepigsgn}\addvspace{\baselineskip}}}
{\ClassError{My Macros}{The first optional argument is not expected in "\epig"}{}}

\newcommand{\itemi}[2][\MyUniPat]{\ifthenelse{\equal{\MyUniPat}{#1}}%
{\begin{itemize}[noitemsep,topsep=3pt] #2 \end{itemize}}%
{\begin{itemize}[#1] #2 \end{itemize}}}

\newcommand{\DoNothing}[1]{}  %

\makeatletter

\protected \def \dg{\@ifstar\dg@st\dg@nost}

\protected \def \dg@nost #1{%
\textcolor{Red}
{
{\normalmarginpar\marginnote{\bl{DG's comment}}}
{\reversemarginpar\marginnote{\bl{DG's comment}}\\}
\IfMathMode{
~~~\txt{#1}~
}{
~\\~~~#1~\\
{\normalmarginpar\marginnote{\bl{\ul{------}}}}
{\reversemarginpar\marginnote{\bl{\ul{------}}}\\}
}
}

}

\protected \def \dg@st #1{%
~\\\textcolor{Red}{%
\reversemarginpar\marginnote{$\blacktriangleleft\blacktriangleright$}%
\normalmarginpar\marginnote{$\blacktriangleleft\blacktriangleright$}%
#1}%
}

\makeatother

\newcommand{\fn}[2][]{%
\IfMathMode{}{}%
\ifthenelse{\equal{}{#1}}%
{\footnote{
\ignorespaces #2}}%
{\footnote{\label{#1}
\ignorespaces #2}}%
}

\makeatletter

\newcommand{\fnref}[1]{%
\protected@xdef\@thefnmark{\ref{#1}}\@footnotemark}

\newcommand{\Nopagebreak}{\par\nobreak\@afterheading} 

\makeatother

\DeclareTextFontCommand{\bemph}{\bfseries}
\DeclareTextFontCommand{\ibemph}{\bfseries\em}

\newcommand{\e}{\emph}

\newcommand{\bl}[1]{{\bf #1}} %

\newcommand{\il}[1]{{\it #1}} %

\def\?{\mskip 1.5mu} %

\newcommand{\tb}{\quad}
\newcommand{\tbb}{\qquad}
\newcommand{\tbbb}{\qquad\qquad}

\newcommand{\MyChangeMathMargins}{%
\setlength{\abovedisplayskip}{\abovedisplayshortskip + 4pt}%
\setlength{\belowdisplayskip}{\abovedisplayshortskip + 5pt}%
}

\newident{\UAMc}{UAM_{compl}}

\newidenT[LAM]{\LAM}{LAM[#1]}

\newidenT[SLAM]{\SLAM}{SLAM[#1]}

\newident[]{\MAp}{MA_{#1}'}
\newident{\MAb}{\wbr{MA}}
\newident{\MAt}{\wtl{MA}}

\newident{\GIPn}{Gut\g-IP_n}

\title{The layer complexity of Arthur-Merlin-like communication}

\newcommand{\instDG}{Institute of Mathematics, Czech Academy of Sciences, \v Zitna 25, Praha 1, Czech Republic.}

\newcommand{\thanksDG}{Partially funded by the grant 19-27871X of GA \v CR.
Part of this work was done while visiting the Centre for Quantum Technologies at the National University of Singapore, and was partially supported by the Singapore National Research Foundation, the Prime Minister's Office and the Ministry of Education under the Research Centres of Excellence programme under grant R 710-000-012-135.}

\author{Dmitry Gavinsky\thanks{\instDG\newline\thanksDG}
}

\begin{document}

\maketitle

\thispagestyle{empty}

\abstart
In communication complexity the \emph{Arthur-Merlin (\AM)} model is the most natural one that allows both randomness and non-determinism.
Presently we do not have any super-logarithmic lower bound for the \AM-complexity of an explicit function.
Obtaining such a bound is a fundamental challenge to our understanding of communication phenomena.

In this %
article  %
we explore the gap between the known techniques and the complexity class \AM.
In the first part we define a new natural class, %
\emph{Small-advantage Layered Arthur-Merlin (\SLAM)},
that %
has  %
the following properties:
\begin{itemize}
\item \SLAM\ is (strictly) included in \AM\ and includes all previously known
subclasses of \AM\  %
with non-trivial lower bounds:
\mal{
\NP, \MA, \SBP, \UAM ~ \sbseq ~ \SLAM ~ \sbs ~ \AM
.}
Note that $\NP\sbs\MA\sbs\SBP$, while \SBP\ and \UAM\ are known to be incomparable.
\item \SLAM\ is qualitatively stronger than the union of those classes:\
\mal{
f\in\SLAM\smin(\SBP\cup\UAM)
}
holds for an (explicit) partial function $f$.
\item \SLAM\ is a subject to the discrepancy bound:\ for any $f$
\mal{
\SLAM(f) \in \Omega\left(\sqrt{\log{\frac1{disc(f)}}}\right)
.}
In particular, the inner product function does not have an efficient \SLAM-protocol.
\end{itemize}
Structurally this can be summarised as
\mal{
\SBP\cup\UAM ~ \sbs ~ \SLAM ~ \sbseq ~ \AM\cap\PP
.}

In the second part we ask why proving a lower bound of $\omega(\sqrt n)$ on the \MA-complexity of an explicit function seems to be difficult.
We show that such a bound either must explore certain ``uniformity'' of \MA\ (which would require a rather unusual argument), or would imply a non-trivial lower bound on the \AM-complexity of the same function.

Both of these results are related to the notion of \emph{layer complexity}, which is, informally, the number of ``layers of non-determinism'' used by a protocol.
\abend

\setcounter{page}{0}
\newpage

\sect[s_intro]{Introduction}

The communication model \emph{Arthur-Merlin (\AM)} is beautiful.
It is the most natural regime that allows both randomness and non-determinism.
Informally,
\itemi{
\item \BPP\ -- the canonical complexity class representing \emph{randomised communication} -- contains such bipartite functions $f$ that admit an \emph{approximate partition} of the set $f^{-1}(1)$ into quasi-polynomially many rectangles;
\item \NP\ -- the canonical complexity class representing \emph{non-deterministic communication} -- contains such $f$ that admit an \emph{exact cover} of the set $f^{-1}(1)$ by quasi-polynomially many rectangles;
\item \AM\ contains such $f$ that admit an \emph{approximate cover} of the set $f^{-1}(1)$ by quasi-polynomially many rectangles.
}

While both \BPP\ and \NP\ are relatively well understood and many strong and tight lower bounds are known, we do not have any non-trivial lower bound for the \AM-complexity of an explicit function.
Obtaining such a bound is a fundamental challenge to our understanding of communication complexity.

Among numerous analytical efforts that have been made to understand \AM, in this %
paper  %
we are paying special attention to these two:
\itemi{
\item In 2003 Klauck~\cite{K03_Rec} studied the class \emph{Merlin-Arthur (\MA)}:\ while (again, informally) \AM\ can be viewed as ``randomness over non-determinism,'' \MA\ is ``non-determinism over randomness.''
Klauck has found an elegant way to %
exploit  %
this difference in order to prove strong lower bounds against \MA.
\item In 2015 G\"o\"os, Pitassi and Watson~\cite{GPW15_Zero} demonstrated strong lower bounds against the class \emph{Unambiguous Arthur-Merlin (\UAM)}, which was defined in the same %
paper.  %
Similarly to \AM\ (and unlike \MA), their class can be viewed as ``randomness over non-determinism,'' but only a very special form of non-determinism is allowed:\ namely, only the (erroneously accepted) elements of $f^{-1}(0)$ may belong to several rectangles; every element of $f^{-1}(1)$ can belong to at most one rectangle of the non-deterministic cover.
In other words, a \UAM-protocol must correspond to an approximate partition of $f^{-1}(1)$, but at the same time it may be an arbitrary cover of a small fraction of $f^{-1}(0)$.
Intuitively, a \UAM-protocol must ``behave like
\BPP\,''
over $f^{-1}(1)$ and is unrestricted over the small erroneously accepted fraction of $f^{-1}(0)$.
}

Interestingly, the classes \MA\ and \UAM\ are \emph{incomparable}:\ from the lower bounds demonstrated in~\cite{GPW15_Zero} and in~\cite{GLMWZ16_Rec}
it follows that   %
\mal{
\UAM \nsbseq \MA
\txt{~~and~~}
\MA \nsbseq \UAM
.}

In the first half of this %
article %
(\sref{s_LAM}) we try to find a communication model that would be as close to \AM\ as possible, while staying within the reach of our analytic abilities.
Inspired by the (somewhat Hegelian) metamorphosis of ``easy'' \BPP\ and \NP\ into ``hard'' \AM, we will try to apply a similar ``fusion'' procedure to the classes \MA\ and \UAM, hoping that the outcome will give us some new insight into the mystery of \AM.

Namely, we start by looking for a communication complexity class, defined as naturally as possible and containing both \MA\ and \UAM.
We will call it \emph{Layered Arthur-Merlin (\LAM)} (Def.~\ref{d_LAM}).
Informally, it can be described as letting a protocol behave like \MA\ over $f^{-1}(1)$ and arbitrarily over the erroneously accepted small fraction of $f^{-1}(0)$.
Note that it follows trivially from the previous discussion (at least on the intuitive level) that
\mal{
\MA\cup\UAM \sbseq \LAM 
.}

Then we will add a few rather technical ``enhancements'' to \LAM\ in order to get a class that includes all previously known classes ``under \AM'' with non-trivial lower bounds:\ most noticeably, the class \SBP, which is known to be strictly stronger than \MA\ and strictly weaker than \AM\ (see~\cite{GPW18_The, GLMWZ16_Rec, K11_On}).

We call the resulting model \emph{Small-advantage Layered Arthur-Merlin (\SLAM)} (Def.~\ref{d_SLAM}) and it holds that
\mal{
\MA, \UAM, \SBP, \LAM ~ \sbseq ~ \SLAM ~ \sbs ~ \AM
.}
Moreover, we will demonstrate a partial function
\mal{
f \in \SLAM \smin (\UAM\cup\SBP)
,}
that is, \SLAM\ will be strictly stronger than the union of all
subclasses of \AM\  %
with previously known non-trivial lower bounds (as $\UAM\cup\SBP$ includes them all).\fn
{
Here and later when referring to the
\emph{subclasses of \AM\    %
with previously known non-trivial lower bounds},'' we mean, in particular, the classes that are \emph{known} to be included by \AM.
Note that not only %
do we  %
not have any non-trivial lower bound against \AM\ yet,  %
but we also cannot guarantee that any interesting complexity class is not included in \AM.
}

Both \LAM\ and \SLAM\ seem to require a new approach for proving lower bounds.
It will be developed in \sref{ss_SLAM_lim}, showing that these classes are still a subject to the discrepancy bound:\ for any function $f$,
\mal{
\SLAM(f) ~\in~ \asOm{\sqrt{\log{\fr1{\disc(f)}}}}
,}
where $\SLAM(f)$ denotes the ``\SLAM-complexity'' of $f$.
In particular, the \emph{inner product function} does not have an efficient \SLAM-protocol.

These properties of \SLAM\ can be summarised structurally as
\mal{
\SBP\cup\UAM ~ \sbs ~ \SLAM ~ \sbseq ~ \AM\cap\PP
,}
where \PP\ is the class consisting of functions with high discrepancy.

The problem of proving a lower bound of $\asom{\sqrt n}$ for the \MA-complexity of an explicit function has been open since 2003, when Klauck~\cite{K03_Rec} showed that the \MA-complexity of \Disj\ and \IP\ was in $\asOm{\sqrt n}$.
At that point a number of researchers believed that the actual \MA-complexity of these problems was in $\asOm n$, so it was surprising when Aaronson and Wigderson~\cite{AW08_Alg} demonstrated \MA-protocols for \Disj\ and \IP\ of cost $\asO{\sqrt n\log n}$, which was later improved by Chen~\cite{C18_On} to $\asO{\sqrt {n\log n\log\log n}}$.

In the second part of this %
article %
(\sref{s_MA}) we try to understand why proving a super-$\sqrt n$ lower bound against \MA\ seems to be difficult.
We will define a communication model \MAt\ that can be viewed -- in certain sense -- as a \emph{non-uniform} \MA.
On the one hand, we will see that imposing the corresponding uniformity constraint on \MAt-protocols makes them not stronger than \MA-protocols; on the other hand, all known lower bounds on $\MA(f)$ readily translate to similar lower bounds on $\MAt(f)$.

Intuitively, a complexity analysis that would explore the uniformity of \MA\ (as opposed to \MAt) must have a very unusual structure:\ the difference between the classes is subtle and we are not aware of any examples where this type of an argument is used.
At the same time, we will see that $\MAt(f) \in \asO{\sqrt{n\tm\AM(f)}}$ for any function $f$ -- that is, any lower bound of the form $\MAt(f) \in \asom{\sqrt n}$ would have non-trivial implications for $\AM(f)$.
This partially explains why no super-$\sqrt n$ lower bound against \MA\ has been found yet.\fn
{
It is relatively easy to show $\AM(f)\in \asOm{\log n}$ for an explicit function (see Footnote~\ref{fn_nlogn}):\ to improve that, a lower bound of the form $\MAt(f) \in \asom{\sqrt{n\tm\log n}}$ would be needed.
However, it seems that proving any $\MAt(f_0) \in \asom{\sqrt n}$ and deriving from it, via the argument of \sref{s_MA}, that $\AM(f_0)\in \asom1$ would by itself shed some new light on the enigma of \AM.
As one of the concluding open problems (Sect.~\ref{s_conc}), we suggest proving a lower bound of the form $\asOm{\sqrt{n\log n}}$ on the \MA-complexity of an explicit function.
}

\paragraph{Why \LAM\ is interesting.}
In the hope that it would benefit the reader, let us explain the motivation for defining and studying the communication models presented in the first part of this %
article.  %
The strong lower bounds that were shown earlier for both \MA\ and \UAM\ were in the first place \emph{steps towards $\AM$}.
Both \MA\ and \UAM\ have very natural definitions, they both can be viewed as weakened versions of \AM, and the authors of both \cite{K03_Rec} and \cite{GPW15_Zero} have invented new insightful approaches while analysing these models.

The model \LAM, in turn, has been defined as a natural ``junction'' of \MA\ and \UAM, at least as strong as either of the predecessors.\fn
{
Here we are referring to \LAM, as its definition is more natural and less technically involved than that of \SLAM; on the other hand, the difference between the two models is, in our opinion, merely formal (as explained above, we wanted the corresponding complexity class to contain all previously studied
subclasses of \AM,  %
including \SBP, and that was the reason for ``boosting'' \LAM, which resulted in \SLAM).
}
As the known approaches for analysing \MA\ and \UAM\ were rather different qualitatively, we expected the new model to be challenging enough to justify defining it.
Our experience of proving a strong lower bound for the newly defined model has confirmed those expectations.

We hope that studying \LAM\ will serve as the next step towards understanding $\AM$.

\sect[s_defi]{Preliminaries and definitions}

For $x\in\OI^n$ and $i\in{[n]}=\set{1\dc n}$, we will write $x_i$ or $x(i)$ to address the
$i$-th bit  %
of $x$ (preferring ``$x_i$'' unless it may cause ambiguity).
Similarly, for $S\sbseq{[n]}$, let both $x_S$ and $x(S)$ denote the $\sz S$-bit string, consisting of (naturally ordered) bits of $x$, whose indices are in $S$.

For a (discrete) set $A$ and $k\in\NN$, we denote by $\Pow(A)$ the set of
subsets of $A$  %
and by $\chs Ak$ the set $\set{a\in\Pow(A)}[\sz a=k]$.

Our primary objects of computation will be \emph{bipartite Boolean functions} of the form $A\times B\to\OI$ (typically, $\OI^n\times\OI^n\to\OI$).
At times we will consider \emph{partial bipartite Boolean functions}, where some of the pairs are \emph{excluded}:\ this can be interpreted either as assuming that those pairs are never given as input, or as allowing any output of a communication protocol when those pairs are received.
We will view partial Boolean functions as total ones that are taking values from $\set{0,1,\bot}$, where ``$\bot$'' marks the excluded input values.
Note that the total functions are a special case, so $f:A\times B\to\set{0,1,\bot}$ can be either total or partial.
When we refer to \emph{an input distribution for a function $f:A\times B\to\set{0,1,\bot}$}, we mean a distribution defined on $f^{-1}(0)\cup f^{-1}(1)$.

We will use the \emph{logical OR $(\bigvee)$} operator with respect to partial Boolean functions, defined as follows (note the asymmetry between the first two cases):
\mal[m_or]{
f_1(x) \vee f_2(y) \deq 
\Cases
{1}{if $f_1(x)=1$ or $f_2(y)=1$;}
{0}{if $f_1(x)=0$ and $f_2(y)=0$;}
{\bot}{otherwise.}
}

\ssect{Communication complexity}

We refer to~\cite{KN97_Comm} for a classical background on communication complexity and to~\cite{GPW18_The} for a great survey of the more recent developments.

\paragraph{Communication problems.}
We will repeatedly consider the following two communication problems.

\ndefi[d_Disj]{Disjointness function, \Disj}{
For every $n\in\NN$, let $(x,y)\in\OI^n\times\OI^n$.
Then
\mal{
\Disj(x,y) = \bigwedge_{i=1}^n (x_i = 0 \vee y_i = 0)
.} 
}

\ndefi[d_IP]{Inner product function, \IP}{
For every $n\in\NN$, let $(x,y)\in\OI^n\times\OI^n$.
Then
\mal{
\IP(x,y) = \bigoplus_{i=1}^n (x_i \wedge y_i)
.} 
}

Both \Disj\ and \IP\ are \emph{total bipartite Boolean functions} -- that is, their input sets are bipartite and the function values are defined for every possible input pair.

\paragraph{Communication models.}
The study of communication complexity was initiated by Abelson~\cite{A78_Lo} in the regime of real-valued messages and adapted by Yao~\cite{Y79_So} to the discrete regime that we are interested in.
The models \Pp\ and \BPP\ that capture one's intuition of \emph{efficient communication} (respectively, deterministic and randomised) date back to~\cite{Y79_So}.
Later Babai, Frankl and Simon~\cite{BFS86_Com} introduced a number of stronger communication models -- in particular, \AM\ and \MA\ -- that intuitively corresponded to some classes studied in the context of structural computational complexity.

\ndefi{Poly-logarithmic, \Pp}{
We call deterministic bipartite communication protocols \emph{\Pp-protocols}.

We denote by \Pp\ the class of functions solved by \Pp-protocols of cost at most $\plog(n)$.
}

\ndefi{Bounded-error Probabilistic Poly-logarithmic, \BPP}{
For every $n\in\NN$, let $f:\OI^n\times\OI^n\to\set{0,1,\bot}$ and $\eps\ge0$.

If for every input distribution $\mu_n$ there exists a \Pp-protocol of cost at most $k_\eps(n)$ that solves $f$ with error at most $\eps$, then we say that the \BPP[\eps]-complexity of $f$, denoted by $\BPP[\eps](f)$, is at most $k_\eps(n)$.

We let the \BPP-complexity of $f$ be its \BPP[\fr13]-complexity.

We denote by \BPP\ the class of functions whose \BPP-complexity is at most $\plog(n)$.
}

The above definition of \BPP\ -- as well those among the following model definitions that are distribution-dependent -- can be phrased in the ``worst-case'' formulations that do not make a reference to input distributions.
Those variants usually correspond to the closures of our definitions with respect to mixed strategies, which, in turn, do not affect the resulting models, due to Von Neumann's minimax principle~\cite{N28_Zur}.

\ndefi{Non-deterministic Poly-logarithmic, \NP}{
For some $k(n)$, let $\Pi=\set{r_i}[i\in[2^{k(n)}]]$ be a family of characteristic functions of combinatorial rectangles over $\OI^n\times\OI^n$.

We call such $\Pi$ an \NP-protocol of cost $k(n)$ that solves the function $f=\bigvee_{i=1}^{2^{k(n)}}r_i(x,y)$ (as well any partial $g$ that is consistent with $f$ on $g^{-1}(0)\cup g^{-1}(1)$).

We say that $\Pi$ accepts every $f^{-1}(1)$ and rejects every $f^{-1}(0)$.

We say that $\Pi$ solves a function $g:\OI^n\times\OI^n\to\set{0,1,\bot}$ with error $\eps$ with respect to an input distribution $\mu_n$, if $\PR[(X,Y)\sim\mu_n]{f(X,Y)\neq g(X,Y)}=\eps$.

We denote by \NP\ the class of functions solved by \NP-protocols of cost at most $\plog(n)$.
}

\ndefi[d_AM]{Arthur-Merlin, \AM}{
For every $n\in\NN$, let $f:\OI^n\times\OI^n\to\set{0,1,\bot}$.

If for every input distribution $\mu_n$ there exists an \NP-protocol of cost at most $k(n)$ that solves $f$ with error at most $\frac13$, then we say that the \AM-complexity of $f$, denoted by $\AM(f)$, is at most $k(n)$.

We denote by \AM\ the class of functions whose \AM-complexity is at most $\plog(n)$.
}

As we mentioned already, \AM\ is a very strong model of communication, for which we currently do not have any non-trivial lower bound.
All the following classes can be viewed (and some of them have been defined) as ``weaker forms'' of \AM: for all of them we already have strong lower bounds.

\ndefi[d_MA]{Merlin-Arthur, \MA}{
For every $n\in\NN$, let $f:\OI^n\times\OI^n\to\set{0,1,\bot}$.

If for some $k(n)$ there are functions $h_1\dc h_{2^{k(n)}}:\OI^n\times\OI^n\to\set{0,1,\bot}$, whose \BPP-complexity is at most $k(n)$, such that $f(x,y)\equiv\bigvee_{i=1}^{2^{k(n)}}h_i(x,y)$, then we say that the \MA-complexity of $f$, denoted by $\MA(f)$, is at most $k(n)$.

We call Merlin-Arthur (\MA) the class of functions whose \MA-complexity is at most $\plog(n)$.
}

Note that ``$\bigvee$'' of partial functions is defined as in~\ref{m_or}.

\ndefi[d_SBP]{Small-advantage Bounded-error Probabilistic Poly-logarithmic, \SBP}{
For every $n\in\NN$, let $f:\OI^n\times\OI^n\to\set{0,1,\bot}$.

If for input distribution $\mu_n$ and some $\alpha>0$ there exists a \Pp-protocol $\Pi$ of cost at most $k'(n)$ such that 
\mal{
&\PR[(X,Y)\sim\mu_n]{\Pi\txt{ accepts }(X,Y)}[{f(X,Y)=1}] \ge \alpha \txt{~~and}\\
&\PR[(X,Y)\sim\mu_n]{\Pi\txt{ accepts }(X,Y)}[{f(X,Y)=0}] \le \fr\alpha2
,}
then we call $\Pi$ an \SBP-protocol for $f$ with respect to $\mu_n$.
The complexity of this protocol is $k'(n)+\log(\frac1\alpha)$ (note that the value of $\alpha$ may depend on both $n$ and $\mu_n$).

If with respect to every $\mu_n$ there exists a \SBP-protocol for $f$ of cost at most $k(n)$, then we say that the \SBP-complexity of $f$, denoted by $\SBP(f)$, is at most $k(n)$.

We denote by \SBP\ the class of functions whose \SBP-complexity is at most $\plog(n)$.
}

It was shown in~\cite{GS86_Pri, BGM06_Err} that $\MA \sbseq \SBP \sbseq \AM$, in~\cite{K11_On} that $\SBP \neq \AM$ and in~\cite{GLMWZ16_Rec} that $\SBP \neq \MA$.
Therefore,
\mal{
\MA \sbs \SBP \sbs \AM 
.~\footnotemark}
\footnotetext
{
Unless stated otherwise, we implicitly assume \emph{partial functions} as the default type of communication problems.
In those cases when the object under consideration is a total function and the fact is significant for the context, that will be mentioned explicitly.
}

The following complexity measure is a core methodological notion for this work.

\ndefi{Layer complexity}{Let $\Pi$ be an \NP-protocol for solving $f:\OI^n\times\OI^n\to\set{0,1,\bot}$, possibly with error.

We say that the protocol $\Pi$\itemi{
\item has layer complexity $l$ if every $(x,y)\in\OI^n\times\OI^n$ belongs to at most $l$ rectangles of $\Pi$;
\item has $0$-layer complexity $l_0$ if every $(x,y)\in f^{-1}(0)$ belongs to at most $l_0$ rectangles of $\Pi$;
\item has $1$-layer complexity $l_1$ if every $(x,y)\in f^{-1}(1)$ belongs to at most $l_1$ rectangles of $\Pi$.
}
}

The concept of layer complexity in the context of non-deterministic communication is very natural and not new, dating back at least to~\cite{KNSW94_Non} by Karchmer, Newman, Saks and Wigderson.
We will use it extensively in order to analyse some previously known
subclasses of \AM\ %
with strong lower bounds and to define some new ones.

The following two classes were introduced quite recently by G\"o\"os, Pitassi and Watson~\cite{GPW15_Zero}.

\ndefi[d_UAM]{Unambiguous Arthur-Merlin, \UAM}{
For every $n\in\NN$, let $f:\OI^n\times\OI^n\to\set{0,1,\bot}$.

If for some constant $\eps<\frac12$ and every input distribution $\mu_n$ there exists an \NP-protocol of cost at most $k(n)$ and $1$-layer complexity $1$ that solves $f$ with error at most $\eps$, then we say that the \UAM-complexity of $f$, denoted by $\UAM(f)$, is at most $k(n)$.

We denote by \UAM\ the class of functions whose \UAM-complexity is at most $\plog(n)$.
}

\ndefi[d_UAMc]{Unambiguous Arthur-Merlin with perfect completeness, \UAMc}{
For every $n\in\NN$, let $f:\OI^n\times\OI^n\to\set{0,1,\bot}$.

If for every input distribution $\mu_n$ there exists an \NP-protocol of cost at most $k(n)$ and $1$-layer complexity $1$ that solves $f$ with perfect completeness (that is, it accepts every $(x,y)\in f^{-1}(1)$) and soundness error at most $\frac12$ (that is, $\PR[\mu_n]{(X,Y)\txt{ is accepted}}[{f(X,Y)=0}]\le\frac12$), then we say that the \UAMc-complexity of $f$, denoted by $\UAMc(f)$, is at most $k(n)$.

We denote by \UAMc\ the class of functions whose \UAMc-complexity is at most $\plog(n)$.
}

The classes \AM, \MA, \UAMc\ and \UAM\ can be defined in an alternative, more ``narrative'' way, where an almighty \emph{prover Merlin} interacts with a limited \emph{verifier Arthur} (who, in turn, is a two-headed union of the \emph{players Alice} and \emph{Bob}).
In the cases of \AM, \UAMc\ and \UAM\ these variants correspond to the closures with respect to mixed strategies (that are equivalent to our definitions, as mentioned earlier).

Note that the error parameter in the definitions of \AM\ and \UAMc\ are fixed without loss of generality, while for \UAM\ it may be any constant $\eps<\frac12$.
In the first two cases the error can be trivially reduced to an arbitrary constant by repeating the protocol constant number of times; on the other hand, in the case of \UAM\ the possibility of efficient error reduction is not known, so fixing a specific $\eps$ might result in weakening the model.\fn
{
In order to reduce the error via repetition, the answer of the new protocol should be the majority vote of the individual answers in the case of \AM, and their logical conjunction in the case of \UAMc.
The problem with this approach for \UAM\ stems from the fact that in order to perform \emph{two-sided} error reduction via repetition, one must take the \emph{majority vote} of the individual answers, which would ruin the required uniqueness of $1$-certificates.
}

It was shown in~\cite{GPW15_Zero} that
$\NP \nsbseq \UAM$.
They also showed that $\UAM \nsbseq \SBP$ held in the context of \emph{query complexity}, later in~\cite{GLMWZ16_Rec} this separation was generalised to the case of communication complexity, thus implying that \UAM\ and \SBP\ are incomparable:
\mal{
\UAM \nsbseq \SBP
\txt{~~and~~}
\SBP \nsbseq \UAM
.}

On the other hand, \UAM\ and \SBP\ are the strongest previously known
communication complexity classes contained in \AM\  %
with non-trivial lower bounds, which makes it interesting to look for their ``natural merge'' and try to prove good lower bounds there.
That will be the quest of the next section.

\sect[s_LAM]{\emph{Layered Arthur-Merlin}:\ getting as close to \AM\ as we can}[Layered Arthur-Merlin:\ getting as close to AM as we can]

Let us try to construct as strong a communication model ``under \AM'' as we can analyse.

We start by considering several slightly stronger modifications of \MA\ that will emphasise the intuition behind the main definitions that will follow.

\ndefi[d_MAp]{\MAp}{
For every $n\in\NN$, let $f:\OI^n\times\OI^n\to\set{0,1,\bot}$.

If for some $k(n)$ and $1\le t(n)\le 2^{k(n)}$ there are functions $h_1\dc h_{t(n)}:\OI^n\times\OI^n\to\set{0,1,\bot}$, whose \BPP[\fr1{3\tm t^2(n)}]-complexity is at most $k(n)$, such that $f(x,y)\equiv\bigvee_{i=1}^{t(n)}h_i(x,y)$, then we say that the \MAp-complexity of $f$, denoted by $\MAp(f)$, is at most $k(n)$.

We call such $\set{h_i}[{i\in[t]}]$ an \MAp-protocol for $f$.
We address the value $t$ as the layer complexity of this protocol.~\fn
{
Note the slight inconsistency in our use of the term \emph{layer complexity} and the parameter $t$ that denotes it:\ most of the time they stand for the actual number of rectangles that an input value belongs to, but occasionally -- in particular, in the context of \MAp\ -- we use them for an \emph{upper bound} on the number of possible ``witnesses.''
The main reason for that is the lack of natural correspondence between the \BPP-protocols (intrinsic to the definition of \MAp) and combinatorial rectangles.
}
}

Observe that 
\mal{
\MA(f)\le\MAp(f)\in\asO{(\MA(f))^2}
}
always holds:\ the inequality follows trivially from the definitions, and the containment results from the well-known fact that for every function $h$ and $\eps>0$, $\BPP[\eps](h)\in\asO{\BPP(h)\tm\log\fr1\eps}$.
So, \MA\ is the class of functions, whose \MAp-complexity is at most $\plog(n)$.

Now suppose $\MAp(f)\le k(n)$, what does it imply with respect to a \emph{known} input distribution $\mu$?
In this case for every $h_i$ there is a \Pp-protocol of cost at most $k(n)$ that computes a function $g_i$, such that $\PR[\mu]{h_i(X,Y)\neq g_i(X,Y)}\le\fr1{3\tm t^2(n)}$; accordingly, the union bound gives
\mal{
\PR[\mu]{f(X,Y)\neq\bigvee_{i=1}^{t(n)}g_i(X,Y)}\le\fr1{3\tm t(n)}
.}

What can we say about a communication complexity class that only requires that the above holds for every $\mu$:\ in particular, what will be its relation to \MA?
Let us define it.

\ndefi[d_MAb]{\MAb}{
For every $n\in\NN$, let $f:\OI^n\times\OI^n\to\set{0,1,\bot}$.

For some $k(n)$ and $1\le t(n)\le 2^{k(n)}$, let $\Pi=\set{g_i}[{i\in[t(n)]}]$ be a family of functions $g_i:\OI^n\times\OI^n\to\set{0,1,\bot}$, whose \Pp-complexity is at most $k(n)$, such that for some input distribution $\mu_n$ it holds that $\PR[\mu_n]{f(X,Y)\neq\bigvee_{i=1}^{t(n)}g_i(X,Y)}\le\fr1{3\tm t(n)}$.
Then we call $\Pi$ an \MAb-protocol of cost $k(n)$ for $f$ with respect to $\mu_n$.

If for every input distribution $\mu_n$ there exists an \MAb-protocol of cost $k(n)$ for $f$, then we say that the \MAb-complexity of $f$, denoted by $\MAb(f)$, is at most $k(n)$.

We denote by \MAb\ the class of functions whose \MAb-complexity is at most $\plog(n)$.
}

Note that
\mal{
\MA\sbseq\MAb
}
follows from the definition and the previous discussion: $\MAb(f)\le\MAp(f)\in\asO{(\MA(f))^2}$.

Towards our goal to construct a communication model under \AM\ as strong as we can analyse, let us look at \UAMc:\ together with \MA\ these are, arguably, the two most natural (though not the strongest) ``sub-\AM'' models for which we have good lower bounds.
Conceptually, the insightful lower bounds given by Klauck~\cite{K03_Rec} for \MA\ and by G\"o\"os, Pitassi and Watson~\cite{GPW15_Zero} for \UAMc\ can be viewed as two different approaches to analysing strong ``sub-\AM'' models of communication complexity.

On the one hand, the more recently defined \UAMc\ has at least one important ``\AM-like'' property that \MA\ lacks:\ \AM\ puts no limitations on the layer complexity of protocols; \MA\ limits the number of ``layers'' over any input pair; \UAMc\ and \UAM\ only limit the $1$-layer complexity (that is, they let the $0$-layer complexity of a protocol be arbitrary, like \AM\ and unlike \MA).
This difference seems to be rather crucial:\itemi{
\item
While the lower-bound argument of~\cite{K03_Rec} against \MA\ can be generalised to work against a communication model that would limit only the $0$-layer complexity of a protocol, it %
does not  %
seem to go through if only the $1$-layer complexity is limited.
\item
If we consider the natural (and the most common) situation when the target function is balanced with respect to its ``hard'' distribution -- which is the case, for instance, for all functions with low discrepancy -- then the ``expected density'' of protocol's rectangles over the points in the (erroneously) accepted $\eps$-fraction of $f^{-1}(0)$ would be much higher than the density in the (rightly) accepted majority of $f^{-1}(1)$.
In other words, the expected number of protocol's rectangles that an \emph{accepted} $(x,y)\in f^{-1}(0)$ belongs to would be considerably higher than the analogous value for $(x,y)\in f^{-1}(1)$.
Accordingly, limiting only the $1$-layer complexity feels like a weaker restriction (i.e., resulting in a stronger defined model) than limiting only the $0$-layer complexity (or both).
}

On the other hand, even though the classes\fn
{
Most of the time the distinction between the two classes is insignificant for the context of this work.
Nevertheless, we will rather often explicitly mention both \UAM\ and \UAMc\ in the same context, as, in our opinion, the former is more naturally defined, while the latter is somewhat simpler to analyse and often allows for clearer intuition.
}
\UAMc\ and \UAM\ limit only the $1$-layer complexity of a protocol, the actual quantitative limitation that they put is way too strong:\ it is $1$, as opposed to the quasi-polynomial limitation on the (total) layer complexity of \MA\ (as emphasised by \defiref{d_MAb}).
For instance, it has been shown in~\cite{GPW15_Zero} that $\NP\nsbseq\UAM$ (note that $\NP\sbseq\MA$ and $\UAMc\sbseq\UAM$).
To include \NP, an ``\NP-like'' class must allow protocols with super-constant $1$-layer complexity.

On the technical level, comparing the definitions of \MAb~(Def.~\ref{d_MAb}) and of \UAMc~(Def.~\ref{d_UAMc}), we can see that in both cases the membership of a function $f$ implies existence of a family of rectangles, whose union approximates $f$ -- that is, existence of good \NP-approximations of $f$:\itemi{
\item if $f\in\MAb$ (in particular, if $f\in\MA$), then for some $t(n)\in\NN$ and every input distribution $\mu_n$ there exists an \NP-protocol of cost at most $\plog(n)$ and layer complexity at most $t(n)$ that solves $f$ with error at most $\fr1{3\tm t(n)}$;
\item if $f\in\UAMc$, then for every input distribution $\mu_n$ there exists an \NP-protocol of cost at most $\plog(n)$ and $1$-layer complexity $1$ that solves $f$ with perfect completeness and soundness error at most $\frac12$ with respect to $\mu_n$.
}
Note that the above membership condition of \UAMc\ is sufficient, and that of \MAb\ is just necessary.

Let us use this intuition to define a new communication complexity class that includes both \UAMc\ and \MAb.

\ndefi[d_LAM]{Layered Arthur-Merlin, \LAM}{
For every $n\in\NN$, let $f:\OI^n\times\OI^n\to\set{0,1,\bot}$.

If for input distribution $\mu_n$ there exists an \NP-protocol $\Pi$ of $1$-layer complexity $t$ that solves $f$ with completeness error at most $\frac13$ and soundness error at most $\frac1{3t}$, then we call $\Pi$ a \LAM-protocol for $f$ with respect to $\mu_n$.
If $\Pi$ contains $K$ rectangles, then its complexity is $\Log{K}$.

If with respect to every $\mu_n$ there exists a \LAM-protocol for $f$ of cost at most $k(n)$, then we say that the \LAM-complexity of $f$, denoted by $\LAM(f)$, is at most $k(n)$.

We denote by \LAM\ the class of functions whose \LAM-complexity is at most $\plog(n)$.
}

It follows readily from the previous discussion that
\mal{
\NP, \MA, \MAb, \UAMc ~ \sbseq ~ \LAM ~ \sbseq ~ \AM
.}

To make it somewhat stronger and to simplify its definition, we have granted to \LAM\ a few additional relaxations (not needed in order to include \MA\ and \UAMc):
Most significantly, in \LAM\ the layer complexity bound $t$ can be chosen per distribution $\mu_n$, and it does not have to be error-independent -- unlike in the cases of both \MA\ and \UAMc\ (for the latter it equals $1$).

Let us further strengthen the model, so that the corresponding complexity class would include all previously known
subclasses of \AM\  %
with strong lower bounds.
The following definition can be viewed as \LAM\ with relaxed accuracy requirements.

\ndefi[d_SLAM]{Small-advantage Layered Arthur-Merlin, \SLAM}{
For every $n\in\NN$, let $f:\OI^n\times\OI^n\to\set{0,1,\bot}$.

If for input distribution $\mu_n$ and some $\alpha>0$ there exists an \NP-protocol $\Pi$ of $1$-layer complexity $t$ such that
\mal{
&\PR[(X,Y)\sim\mu_n]{\Pi\txt{ accepts }(X,Y)}[{f(X,Y)=1}] \ge \alpha \txt{~~and}\\
&\PR[(X,Y)\sim\mu_n]{\Pi\txt{ accepts }(X,Y)}[{f(X,Y)=0}]\le\fr\alpha{2t}
,}
then we call $\Pi$ a \SLAM-protocol for $f$ with respect to $\mu_n$.
If $\Pi$ contains $K$ rectangles, then its complexity is $\Log{\fr K\alpha}$ (the value of $\alpha$ may depend on both $n$ and $\mu_n$).

If with respect to every $\mu_n$ there exists a \SLAM-protocol for $f$ of cost at most $k(n)$, then we say that the \SLAM-complexity of $f$, denoted by $\SLAM(f)$, is at most $k(n)$.

We denote by \SLAM\ the class of functions whose \SLAM-complexity is at most $\plog(n)$.
}

As any \LAM-protocol of cost $k$ is also a \SLAM-protocol of cost $k+\log\fr32$,
\mal{
\SLAM(f) < \LAM(f) + 1
}
holds for all $f$.

Later (\sref{ss_SBP-UAM}) we will see that \SLAM\ indeed is a proper
subclass of \AM\  %
that includes all previously known (as far as we are aware)
subclasses of \AM\ %
with strong lower bounds:
\mal{
\NP, \MA, \MAb, \UAMc, \LAM, \UAM, \SBP ~ \sbseq ~ \SLAM ~ \sbs ~ \AM
;}
moreover, it is strictly stronger than their union:
\mal{
\UAM\cup\SBP \sbs \SLAM
.}

\ssect[ss_SLAM_lim]{Limitations of \LAM\ and \SLAM}[Limitations of LAM and SLAM]

Let us see that the \SLAM-complexity is a subject to the discrepancy bound.

\ndefi[d_disc]{Discrepancy}{
For every $n\in\NN$, let $f:\OI^n\times\OI^n\to\set{0,1,\bot}$ and $\mu_n$ be a distribution on $\OI^n\times\OI^n$.

The discrepancy of $f$ with respect to $\mu_n$ is defined as
\mal{
\disc[\mu_n](f) = \Max{r}[{\sz{\mu_n(r\cap f^{-1}(1)) - \mu_n(r\cap f^{-1}(0))}}]
,}
where $r$ ranges over the combinatorial rectangles over $\OI^n\times\OI^n$.

We denote $\disc(f) = \Min{\mu}[{\disc[\mu](f)}]$.
}

\theo[t_LAM]{For any $f:\OI^n\times\OI^n\to\set{0,1,\bot}$:
\mal{
\SLAM(f) ~\in~ \asOm{\sqrt{\log{\fr1{\disc(f)}}}}
.}
}

That is,
\mal{
\SLAM ~ \sbseq ~ \PP
,}
where \PP\ is the class consisting of functions with high discrepancy.
Along with other mentioned properties, this implies
\mal{
\SBP\cup\UAM ~ \sbs ~ \SLAM ~ \sbseq ~ \AM\cap\PP
\txt{~~and~~}
\SLAM ~ \sbs ~ \AM
,}
as $\AM\nsbseq\PP$ is known~\cite{K11_On}.

\crl[cr_IP]{Any \LAM- or \SLAM-protocol for \IP\ has cost $\asOm{\sqrt n}$.}

To prove the theorem we will use the following combinatorial lemma.

\lem[l_sets]{Let $C_1\dc C_m$ be finite sets and
$W\deq \bigcup_{i=1}^m C_i$. %
Let $t\in\NN$, $\beta>1$ and
\mal{
& W_0 \deq \set{w\in W}[{1 \le |\set{i\in[m]}[w\in C_i]| \le t}],\\
& W_1 \deq \set{w\in W}[{|\set{i\in[m]}[w\in C_i]| \ge \beta\tm t}]
.}
Let $\mu$ be a distribution on $W$, such that
\mal{
\fr{\mu(W_1)}{\mu(W_0)} \ge \lambda
}
for some $\lambda>0$.~\fn[fn_div0]
{
Here let $\fr x0>y$ hold for any $x,y>0$.
}

Then for any $\gamma>\lambda$ there exists $J\sbseq[m]$ of size at most $k\deq\ceil{\Log[\fr{\beta+1}2]{\fr\gamma\lambda}}$, such that for
$C_J\deq\bigcap_{j\in J}C_j$  %
it holds that
\mal[m_g_req]{
\fr{\mu(C_J \cap W_1)}{\mu(C_J \cap W_0)} \ge \gamma,~\fnref{fn_div0}
}
and
\mal{
\mu(C_J \cap W_1) \ge \mu(W_1) \tm \left( \fr{\Min{1,\, \beta-1}}{2m} \right)^k
.}
}

Informally, the lemma says that for any family of sets $C_1\dc C_m$ there exists
$C_J=\bigcap_{j\in J}C_j$  %
that ``highlights'' points that are contained in more than a certain threshold number of %
sets $C_i$;  %
moreover, the fraction of such points in $W$ that end up in $C_J \cap W$ is not too small.

\prfstart[\lemref{l_sets}]
We will find such $C_{i_0}$ that $\mu(C_{i_0} \cap W_1)$ is not too small and, at the same time, $\fr{\mu(C_{i_0} \cap W_1)}{\mu(C_{i_0} \cap W_0)}$ is significantly larger than $\fr{\mu(W_1)}{\mu(W_0)}$.
The result will follow by induction on $t$.

The first part of the argument is the same for the base case ($t=1$) and the inductive step ($t\ge2$).
Let $C_i(\dt)$ denote the characteristic function of $C_i$, then
\mal[P]{
& \E[X\sim\mu]{\sum_{i=1}^m C_i(X)}[{X\in W_1}]
\ge \beta \tm \E[X\sim\mu]{\sum_{i=1}^m C_i(X)}[{X\in W_0}];\\
& \sum_{i=1}^m \E[X\sim\mu]{C_i(X)}[{X\in W_1}]
\ge \beta \tm \sum_{i=1}^m \E[X\sim\mu]{C_i(X)}[{X\in W_0}];\\
& \sum_{i=1}^m
\left(
\E[X\sim\mu]{C_i(X)}[{X\in W_1}]
- \beta \tm \E[X\sim\mu]{C_i(X)}[{X\in W_0}]
\right)
\ge 0;\\
0 ~ \le ~
& \sum_{i=1}^m
\left(
\fr{\beta+1}{2\beta} \tm \E[X\sim\mu]{C_i(X)}[{X\in W_1}]
- \fr{\beta+1}2 \tm \E[X\sim\mu]{C_i(X)}[{X\in W_0}]
\right)\\
&\tbb = \sum_{i=1}^m
\bigg(
\E[X\sim\mu]{C_i(X)}[{X\in W_1}]
- \fr{\beta-1}{2\beta} \tm \E[X\sim\mu]{C_i(X)}[{X\in W_1}]\\
&\tbbb - \fr{\beta+1}2 \tm \E[X\sim\mu]{C_i(X)}[{X\in W_0}]
\bigg);\\
& \sum_{i=1}^m
\left(
\E[X\sim\mu]{C_i(X)}[{X\in W_1}]
- \fr{\beta+1}2 \tm \E[X\sim\mu]{C_i(X)}[{X\in W_0}]
\right)\\
&\tbb \ge \fr{\beta-1}{2\beta}
\tm \E[X\sim\mu]{\sum_{i=1}^m C_i(X)}[{X\in W_1}]\\
&\tbb \ge \fr{\beta-1}{2\beta} \tm \beta \tm t
~ \ge ~ \fr{\beta-1}2;\\
\exists i_0\in[m]: ~
& \E[X\sim\mu]{C_{i_0}(X)}[{X\in W_1}]
- \fr{\beta+1}2 \tm \E[X\sim\mu]{C_{i_0}(X)}[{X\in W_0}]
~ \ge ~ \fr{\beta-1}{2m}
.}
That is,
\mal{
\fr{\mu(C_{i_0} \cap W_1)}{\mu(W_1)}
- \fr{\beta+1}2 \tm \fr{\mu(C_{i_0} \cap W_0)}{\mu(W_0)}
\ge \fr{\beta-1}{2m}
,}
which implies
\mal[m_C-size]{
\mu(C_{i_0} \cap W_1) \ge \fr{\beta-1}{2m} \tm \mu(W_1)
}
and
\mal[m_C-boost]{
\fr{\mu(C_{i_0} \cap W_1)}{\mu(C_{i_0} \cap W_0)}
\ge \fr{\beta+1}2 \tm \fr{\mu(W_1)}{\mu(W_0)}
\ge \fr{\beta+1}2 \tm \lambda
.}

At this point we check whether letting $J=\set{i_0}$ would satisfy the statement of the lemma.
Assume that it %
would not;   %
as $\gamma>\lambda \To k\ge1$, this necessarily means that \bref{m_C-boost} is insufficient to guarantee \bref{m_g_req}.
In other words, it holds that
\mal{
\fr{\beta+1}2 \tm \lambda
\le \fr{\mu(C_{i_0} \cap W_1)}{\mu(C_{i_0} \cap W_0)}
< \gamma
,}
where the latter inequality is the contrapositive of \bref{m_g_req} with respect to $J=\set{i_0}$, and therefore
\mal[m_to-cont]{
\fr{\beta+1}2 < \fr\gamma\lambda
.}

Denote
\mal{
C_j' \deq C_j \cap C_{i_0}
}
for all $j\neq i_0$ and
\mal{
C_{i_0}'' \deq C_{i_0} \smin
\bigcup_{j\neq i_0}C_j %
.}
Note that $C_{i_0}'' \sbseq W_0$.

How we continue from here depends on the value of $t$:\ first suppose that $t=1$ (the base case for the induction).
Let $i_1\in[m]\mset{i_0}$ be such that $\mu(C_{i_1}' \cap W_1)$ is maximised.
Let $J \deq \set{i_0, i_1}$.
From \bref{m_to-cont} it follows that
\mal{
k = \ceil{\Log[\fr{\beta+1}2]{\fr\gamma\lambda}} \ge 2 = \sz{J}
.}
From \bref{m_C-size} and the choice of $i_1$,
\mal{
\mu(C_J \cap W_1)
\ge \fr1{m-1} \tm \fr{\beta-1}{2m} \tm \mu(W_1)
> \mu(W_1) \tm \left( \fr{\Min{1,\, \beta-1}}{2m} \right)^2
.}
As $t=1$,
\mal{
\mu(C_J \cap W_0) =0
,}
which satisfies \bref{m_g_req} (according to Footnote~\ref{fn_div0}).
This finishes the proof of the base case.

Now suppose $t \ge 2$.
That is, we are inside the inductive step, so let us apply \lemref{l_sets} inductively to the family $\set{C_j'}[j\neq i_0]$ with parameters
\mal{
m' = m-1,~ t' = t-1,~ \beta' = \fr{\beta\tm t-1}{t-1},~ \gamma' = \gamma
.}
Note that this choice corresponds to
\mal{
& W' = ( W \cap C_{i_0} ) \smin C_{i_0}'',~
W_0' = ( W_0 \cap C_{i_0} ) \smin C_{i_0}'',~
W_1' = W_1 \cap C_{i_0},\\
& \lambda' = \fr{\beta+1}2 \tm \lambda
,}
where the last equality follows from \bref{m_C-boost}.
Note that $\lambda'<\gamma'$ follows from \bref{m_to-cont}.

The lemma guarantees existence of (non-empty) $J'\sbseq [m]\mset{i_0}$ of size at most
\mal{
k' = \ceil{\Log[\fr{\beta'+1}2]{\fr\gamma{\lambda'}}}
\le \ceil{\Log[\fr{\beta+1}2]{\fr\gamma{\lambda'}}}
= \ceil{\Log[\fr{\beta+1}2]{\fr\gamma\lambda} - 1}
= k-1
}
(where the inequality follows from $\beta' > \beta$), such that for
$C_{J'} \deq \bigcap_{j\in J'}C_j'$  %
it holds that
\mal{
\fr{\mu(C_{J'} \cap C_{i_0} \cap W_1)}{\mu(C_{J'} \cap C_{i_0} \cap W_0)}
& \ge \fr{\mu(C_{J'} \cap W_1')}{\mu(C_{J'} \cap W_0')}
\ge \gamma;\\
\mu(C_{J'} \cap C_{i_0} \cap W_1)
& = \mu(C_{J'} \cap W_1')
\ge \mu(W_1') \tm \left( \fr{\Min{1,\, \beta-1}}{2m} \right)^{k'}\\
& \ge \mu(C_{i_0} \cap W_1) \tm \left( \fr{\Min{1,\, \beta-1}}{2m} \right)^{k-1}
\ge \mu(W_1) \tm \left( \fr{\Min{1,\, \beta-1}}{2m} \right)^k
,}
where the last inequality follows from \bref{m_C-size}.
Letting $J \deq J' \cup \set{i_0}$ finishes the proof.
\prfend

We are ready to prove the lower bound.

\prfstart[\theoref{t_LAM}]
The argument is as follows.
Recall that the core advantage of the models \LAM\ and \SLAM\ over \MA\ is allowing arbitrarily high $0$-layer complexity in efficient protocols:
If the $0$-layer complexity of a given protocol $\Pi$ is not above its $1$-layer complexity, then Klauck's argument for \MA\ limits $\Pi$'s strength.

The case when the $0$-layer complexity is high was the main challenge of this work and the reason why it was written.
Here we further assume that the \emph{average} $0$-layer complexity of $\Pi$ is \emph{noticeably} higher than its \emph{average} $1$-layer complexity (we notice that the other cases can be handled by a straightforward adaptation of Klauck's argument).
The layer complexity measures the density of $\Pi$'s rectangles, and the assumed difference in the average densities between $f^{-1}(0)$ and $f^{-1}(1)$ implies that the membership of a pair of random variables $(X, Y)$ in ``too many'' rectangles makes the event $[f(X, Y)=0]$ more likely.
\lemref{l_sets} gives us a ``not too small'' rectangle intersection -- therefore a rectangle by itself -- where many elements belong to many (original) rectangles. 
The discrepancy assumption applied to this new rectangle concludes the argument.

Let $\mu$ be a distribution that achieves $\disc(f) = \disc[\mu](f)$ and
\mal{
\Pi = \set{r_i}[{i\in[2^{k(n)}]}]
}
be a \SLAM-protocol for $f$ with respect to $\mu$ of cost $k(n)+\log(\fr1{\alpha(n)})$ that accepts $\alpha(n)$-fraction of the elements of $f^{-1}(1)$, and whose $1$-layer complexity is $t(n)$.

By the definition of \SLAM, the soundness error of $\Pi$ in solving $f$ with respect to $\mu$ is at most
\mal[m_es]{
\fr{\alpha(n)}{2\tm t(n)}
.}

By the definition of $\disc[\mu]$ (and the fact that $\OI^n\times\OI^n$ is a rectangle),
\mal[m_bala]{
\PR[(X,Y)\sim\mu]{f(X,Y)=0}, \PR[(X,Y)\sim\mu]{f(X,Y)=1}
~ \in ~
\fr12 \pm \fr{\disc[\mu](f)}2 = \fr12 \pm \fr{\disc(f)}2
.}

Let $l_0^{av}(n)$ denote the \emph{average} $0$-layer complexity of $\Pi$, namely
\mal{
l_0^{av}(n) \deq \E[(X,Y)\sim\mu]{|\set{r\in\Pi}[(X,Y)\in r]|}[{(X,Y)\in\Pi^{-1}(1) \cap f^{-1}(0)}]
,}
where ``$\Pi^{-1}(1)$'' denotes the set of input pairs accepted by $\Pi$.

Fix $n\in\NN$.
We will consider two cases, distinguished by the value of $l_0^{av}(n)$.

First suppose that
\mal[m_case1]{
l_0^{av}(n) \le \fr{4\tm t(n)}3
.}
Then
\mal{
\sum_{r\in\Pi}\mu(r\cap f^{-1}(1))
& \ge \PR[(X,Y)\sim\mu]{(X,Y)\in\Pi^{-1}(1) \cap f^{-1}(1)}\\
& = \PR[\mu]{f(X,Y)=1}
\tm \PR[\mu]{(X,Y)\in\Pi^{-1}(1)}[{f(X,Y)=1}]\\
& \ge \left( \fr12 - \fr{\disc(f)}2 \right) \tm \alpha(n)
,}
where the last inequality follows from \bref{m_bala}, and
\mal[P]{
\sum_{r\in\Pi}\mu(r\cap f^{-1}(0))
& = \sum_{r\in\Pi} \sum_{(x,y)\in r\cap f^{-1}(0)} \mu(x,y)\\
& = \sum_{(x,y)\in f^{-1}(0)} \mu(x,y)\tm\sz{\set{r\in\Pi}[(x,y)\in r]}\\
& = \PR[(X,Y)\sim\mu]{f(X,Y)=0}
\tm \E[(X,Y)\sim\mu]{|\set{r\in\Pi}[(X,Y)\in r]|}[{f(X,Y)=0}]\\
& = \PR[\mu]{(X,Y)\in\Pi^{-1}(1) \cap f^{-1}(0)}\\
&\tbb \tm \E[\mu]{|\set{r\in\Pi}[(X,Y)\in r]|}
[{(X,Y)\in\Pi^{-1}(1) \cap f^{-1}(0)}]\\
& = \PR[\mu]{(X,Y)\in\Pi^{-1}(1) \cap f^{-1}(0)}
\tm l_0^{av}(n)\\
& \le \PR[\mu]{f(X,Y)=0}
\tm \fr{\alpha(n)}{2\tm t(n)}
\tm \fr{4\tm t(n)}3\\
& \le \left( \fr12 + \fr{\disc(f)}2 \right) \tm \fr{2\tm \alpha(n)}3
,}
where the first inequality follows from \bref{m_es} and \bref{m_case1}, and the last one from \bref{m_bala}.
Therefore,
\mal{
\sum_{r\in\Pi} \mu(r\cap f^{-1}(1)) - \mu(r\cap f^{-1}(0))
& \ge \left( \fr16 - \fr{5\tm\disc(f)}6 \right) \tm \alpha(n)
}
and for some $r_0\in\Pi$ it holds that
\mal{
\disc(f)
\ge \mu(r_0\cap f^{-1}(1)) - \mu(r_0\cap f^{-1}(0))
\ge \left( \fr16 - \fr{5\tm\disc(f)}6 \right) \tm \alpha(n) \tm 2^{-k(n)}
}
and
\mal{
k(n) + \Log{\fr1{\alpha(n)}} \in \Log{\fr1{\disc(f)}} - \asO1
,}
as required.

Now suppose that
\mal{
l_0^{av}(n) > \fr{4\tm t(n)}3
.}

Define
\mal{
A \deq \set{(x,y)}[{\sz{\set{r\in\Pi}[(x,y)\in r]} \ge \fr{5\tm t(n)}4}]
.}
Let us see that $\mu(A)$ cannot be too small.
\mal{
\fr{4\tm t(n)}3
& < l_0^{av}(n)
= \E[\mu]{|\set{r\in\Pi}[(X,Y)\in r]|}[{(X,Y)\in\Pi^{-1}(1) \cap f^{-1}(0)}]\\
& \le \PR[\mu]{(X,Y)\in A}[{(X,Y)\in\Pi^{-1}(1) \cap f^{-1}(0)}]
\tm 2^{k(n)}\\
& \tb + \left(1 - \PR[\mu]{(X,Y)\in A}[{(X,Y)\in\Pi^{-1}(1) \cap f^{-1}(0)}] \right)
\tm \fr{5\tm t(n)}4\\
& \le \fr{5\tm t(n)}4
+ 2^{k(n)} \tm \PR[\mu]{(X,Y)\in A}[{(X,Y)\in\Pi^{-1}(1) \cap f^{-1}(0)}]
.}
Therefore,
\mal{
\PR[\mu]{(X,Y)\in A}[{(X,Y)\in\Pi^{-1}(1) \cap f^{-1}(0)}]
> \fr{t(n)\tm2^{-k(n)}}{12}
}
and
\mal{
\PR[\mu]{(X,Y)\in A}
> \fr{\mu\left(\Pi^{-1}(1) \cap f^{-1}(0)\right)}{12\tm2^{k(n)}}
.}
On the other hand,
\mal{
\mu\left(\Pi^{-1}(1) \cap f^{-1}(0)\right)
\ge \Max{r\in\Pi}[{\mu\left(r\cap f^{-1}(0)\right)}]
\ge \fr{\alpha(n)}{2^{k(n)}} \tm \left( \fr12 - \fr{\disc(f)}2 \right)^2
,}
where the last inequality follows from the fact that the $\mu$-weight of the largest rectangle of $\Pi$ is, due to~\bref{m_bala}, at least $\fr{\alpha(n)}{2^{k(n)}} \tm (\fr12 - \fr{\disc(f)}2)$, and the relative $\mu$-weight of $f^{-1}(0)$ in it is at least $\fr12 - \fr{\disc(f)}2$.
Assuming $[\disc(f)\le\frac12]$ (otherwise the desired statement holds trivially), we get
\mal[m_sA]{
\mu(A) \ge \fr{\alpha(n)}{192\tm2^{2\tm k(n)}}
~.}

Let
\mal{
B \deq \set{(x,y)}[1 \le |\set{r\in\Pi}[(x,y)\in r]| \le t(n)]
,}
then
\mal[m_sB]{
\mu(B) \le \mu\left(\Pi^{-1}(1)\right) < \alpha(n)
,}
as $\Pi$ accepts, with respect to $\mu$, $\alpha(n)$-fraction of $f^{-1}(1)$ and smaller fraction of $f^{-1}(0)$.

Note that
\mal[m_AB]{
A\sbs f^{-1}(0)
\txt{~~and~~}
f^{-1}(1) \cap \Pi^{-1}(1) \sbseq B
,}
as follows from their definitions and the fact that the $1$-layer complexity of $\Pi$ is $t(n)$.

Let us make use of the difference in the ``rectangle density'' of $A$ and $B$ via applying \lemref{l_sets}.
Namely, let $m\deq2^{k(n)}$, $C_i\deq r_i$, $t\deq t(n)$, $\beta\deq\fr54$ and $\gamma\deq2$. 
Then the conditions of the lemma hold with respect to $W_0=B$, $W_1=A$ and $\lambda=\fr1{192\tm2^{2\tm k(n)}}$.
Then there exists some $J\sbseq[2^{k(n)}]$, such that for
\mal{
s\deq\bigcap_{j\in J}r_j
}
it holds that
\mal{
\fr{\mu(s \cap A)}{\mu(s \cap B)} \ge 2
}
and
\mal{
\mu(s \cap A) \in \alpha(n) \tm 2^{-\asO{k^2(n)}}
,}
as follows from \bref{m_sA}, \bref{m_sB} and the statement of the lemma.

That is,
\mal{
\mu(s \cap A) - \mu(s \cap B)
\ge \fr{\mu(s \cap A)}2
\in \alpha(n) \tm 2^{-\asO{k^2(n)}}
.}
As $s\sbseq\Pi^{-1}(1)$, it follows from \bref{m_AB} that
\mal{
\mu(s \cap f^{-1}(0)) - \mu(s \cap f^{-1}(1))
\in \alpha(n) \tm 2^{-\asO{k^2(n)}}
,}
and since $s$ is a rectangles' intersection and therefore a rectangle itself,
\mal{
\disc(f) \in \alpha(n) \tm 2^{-\asO{k^2(n)}}
,}
that is,
\mal{
\asO{k^2(n)} + \Log{\fr1{\alpha(n)}} \ge \Log{\fr1{\disc(f)}}
~~ \To ~~
k(n) + \Log{\fr1{\alpha(n)}} \in \asOm{\sqrt{\log{\fr1{\disc(f)}}}}
,}
as required.
\prfend

\ssect[ss_SBP-UAM]{More about \LAM\ and \SLAM}[More about LAM and SLAM]

When we defined the communication complexity class \SLAM\ (Def.~\ref{d_SLAM}), we promised to show later that
\mal{
\NP, \MA, \UAMc, \UAM, \SBP, \LAM ~ \sbseq ~ \SLAM ~ \sbs ~ \AM
}
and
\mal{
\UAM\cup\SBP \sbs \SLAM
.}

The relations
\mal{
\NP, \MA, \UAMc \sbseq \LAM \sbseq \AM
\txt{~~and~~}
\LAM  \sbseq  \SLAM
}
follow trivially from the definitions.
\theoref{t_LAM} implies that
\mal{
\SLAM ~ \sbseq ~ \PP
~~ \Then ~~
\SLAM ~ \sbs ~ \AM
,}
as \PP\ is the class consisting of functions with high discrepancy and $\AM\nsbseq\PP$ is known~\cite{K11_On}.

It remains to see that 
\mal{
\UAM\cup\SBP \sbs \SLAM \sbseq ~ \AM
,}
which will be implied by the upcoming \clmref[c_SLAM-AM]{c_UAM-SBP-SLAM}.

\clm[c_SLAM-AM]{For any bipartite Boolean function $f$:
\mal{
\AM(f) ~\in~ \asO{\SLAM(f) + \log n}
.}
}

\prfstart[\clmref{c_SLAM-AM}]
The proof combines the ``randomness sparsification'' method of Goldwasser and Sipser~\cite{GS86_Pri} with \NP-witnessing.

Assume $\SLAM(f)=k(n)$.
Then by Von Neumann's minimax principle~\cite{N28_Zur} there exists a family
$\Pi=\set{h_1\dc h_m}$ for some $m\in\NN$, where every $h_i$ is a bipartite Boolean function computable by an \NP-protocol of cost at most $k(n)$, such that
\mal{
\forall (x,y):~\sz{\set{i\in[m]}[h_i(x,y)=1]}
~\Cases
{\ge \alpha\tm m}{if $f(x,y)=1$,}
{\le \fr\alpha2\tm m}{if $f(x,y)=0$}
}
for some $\alpha\ge2^{-k(n)}$.\fn
{
Note that the actual value of $t$ is insignificant for this argument, which is not surprising:\ if we modify the definition of \SLAM\ (Def.~\ref{d_SLAM}) by allowing arbitrary $1$-layer complexity, then we end up with a definition of \AM.
}

By a standard Bernstein-type concentration argument (e.g.,~\cite{DM05_Con}, Lemma~1, \emph{Hoeffding inequality}),
there exists $l\in\asO{\frac n\alpha}\sbseq\asO{2^{k(n)+\log n}}$ such that for some $\Pi'\sbseq\Pi$ of size $l$ it holds that
\mal{
\forall (x,y):~\sz{\set{i\in[l]}[h_i(x,y)=1]}
~\Cases
{\ge \fr{2\alpha}3\tm l}{if $f(x,y)=1$,}
{\le \fr\alpha3\tm l}{if $f(x,y)=0$,}
}
where we have assumed without loss of generality that $\Pi'=\set{h_1\dc h_l}$.

By another application of the Hoeffding inequality, %
for some $s\in\asO{n+ \frac1\alpha}\sbseq\asO{2^{k(n)+\log n}}$ and a uniformly random function $g:[l]\to[s]$ it holds with positive probability that
\mal[m_sparse]{
\forall (x,y):
~\PR[{Z\unin[s]}]
{\exists i\in[l]:~ h_i(x,y)=1 \wedge g(i) = Z}
~\Cases
{\ge \fr35}{if $f(x,y)=1$,}
{\le \fr25}{if $f(x,y)=0$.}
}
Fix any such $g$.

Consider the following \AM-protocol (described below in a distribution-free regime, which is the dual equivalent of \defiref{d_AM}).
\itemi{
\item The players pick $Z\unin[s]$ and send it to Merlin.
\item Merlin responds with $i\in[l]$ and $w\in\OI^{k(n)}$.
\item The players accept if and only if $h_i(X,Y)=1 \wedge g(i) = Z$, where the former is witnessed by $w$ (recall that $\NP(h_i)\le k(n)$).
}

By \bref{m_sparse}, this is an \AM-protocol for $f$ with error at most $\fr25$; repeating it several times and taking the majority vote brings the error bound to at most $\fr13$.
The cost of the resulting protocol is in $\asO{k(n)+\log n}$, as required.
\prfend

To see that $\UAM\cup\SBP \sbs \SLAM$, we prove a somewhat stronger separation:
\mal{
\LAM \nsbseq \UAM\cup\SBP
.}
For that we will use several results from~\cite{GPW15_Zero, GLMWZ16_Rec}.

\nfct[f_Disj]{$\NP \nsbseq \UAM$~\cite{GPW15_Zero}}{
Let $\lnot\Disj(x,y)\deq1-\Disj(x,y)$ for every $(x,y)\in\OI^n\times\OI^n$, then
\mal{
\UAM(\lnot\Disj) \in \asOm n
.}
}

The following partial function has been used to show that $\UAMc \nsbseq \SBP$.

\ndefi[d_GIP]{\GIPn~\cite{GPW15_Zero, GLMWZ16_Rec}}{
For an even $m\in\NN$, let $n=m^2\tm\ceil{200\tm\log m}$.
For any $x\in\OI^n$ and $i,j\in[m]$, let $x_{i,j}$ denote the sub-string of $x$ that starts from bit $\ceil{200\tm\log m}\tm(m\tm(i-1)+j-1)+1$ and contains $\ceil{200\tm\log m}$ bits.
Denote
\mal{
\forall i\in[m]:~\sharp_i \deq \sz{\set{j}[\IP(x_{i,j},y_{i,j})=1]}
,}
then
\mal{
\GIPn(x,y) = \Cases
{1}{if $\forall i:\sharp_i=1$;}
{0}{if $\sz{\set{i}[\sharp_i=0]}=\sz{\set{i}[\sharp_i=2]}=\fr m2$;}
{\bot}{otherwise.}
}
}

\nfct[f_GIP]{$\UAMc \nsbseq \SBP$~\cite{GPW15_Zero, GLMWZ16_Rec}}{
\mal{
& \UAMc(\GIPn) \in \asO{\log n};\\
& \SBP(\GIPn) \in \asOm{\sqrt m\tm\log m} = \asOm{n^{\frac14}\tm\log^{\frac34}n}
.}
}

\clm[c_UAM-SBP-SLAM]{For any $n\in\NN$ such that $\GIPn$ is defined and
$x_1,x_2,y_1,y_2\in\OI^n$, let
\mal{
f((x_1,x_2),(y_1,y_2)) \deq \lnot\Disj(x_1,y_1) \wedge \GIPn(x_2,y_2)
.}
Then
\mal{
& \LAM(f),\,\SLAM(f) \in \asO{\log^2 n};\\
& \UAM(f) \in \asOm n;\\
& \SBP(f) \in \asOm{n^{\frac14}\tm\log^{\frac34}n}
.}
}

\prfstart[\clmref{c_UAM-SBP-SLAM}]
Consider an input distribution $\mu$ that fixes $X_1=Y_1=1^n$ and makes the pair $(X_2,Y_2)$ come from a hard distribution for $\GIPn(X_2,Y_2)$, then any \SBP-protocol that solves $f$ with respect to $\mu$ must have complexity $\asOm{n^{\frac14}\tm\log^{\frac34}n}$, according to \fctref{f_GIP}.
Similarly, a distribution that fixes $(X_2,Y_2)\in\GIPn^{-1}(1)$ arbitrarily and makes $\Disj(X_1,Y_1)$ hard for \UAM\ witnesses that $\UAM(f) \in \asOm n$, according to \fctref{f_Disj}.

To see that $\LAM(f) \in \asO{\log^2 n}$, let $\mu$ be any input distribution for $f$ and let $\mu'$ be the marginal distribution of $(X_2,Y_2)$ when $((X_1,X_2),(Y_1,Y_2))\sim\mu$.
Consider a \UAMc-protocol $\Pi$ of complexity $\asO{\log n}$ that solves $\GIPn$ with perfect completeness and soundness error at most $\frac12$ with respect to $\mu'$, and let $\Pi'$ be its amplified version of complexity $\asO{\log^2 n}$ that solves $\GIPn$ with soundness error at most $\frac1{3n}$ with respect to $\mu'$.

Let $\Pi''((X_1,X_2),(Y_1,Y_2))$ be a non-deterministic protocol for $f$ that does the following:
\itemi{
\item emulates the behaviour of $\Pi'(X_2,Y_2)$;
\item runs the trivial \NP-protocol for $\lnot\Disj(X_1,Y_1)$;
\item accepts if and only if the two steps above have accepted.
}
The complexity of $\Pi''$ is in $\asO{\log^2 n}$.

Since an \NP-protocol for $\lnot\Disj$ is exact (though non-deterministic), an error can come only from the first step; since $\Pi'$ has perfect completeness, so does $\Pi''$.
The soundness error of $\Pi''$ in solving $f$ with respect to $\mu$ equals that of $\Pi'$ in solving $\GIPn$ with respect to $\mu'$, which is at most $\frac1{3n}$.
Since $\Pi'$ has $1$-layer complexity $1$, the $1$-layer complexity of $\Pi''$ equals that of the \NP-protocol for $\lnot\Disj$, which is $n$.
So, $\Pi''$ is a valid \LAM-protocol for $f$ with respect to $\mu$, as required.
\prfend

\sect[s_MA]{On proving super-$\sqrt n$ lower bounds against \MA}[On proving super-square-root lower bounds against MA]

When Klauck~\cite{K03_Rec} showed that $\MA(\Disj), \MA(\IP)\in\asOm{\sqrt n}$, many believed that the actual \MA-complexity of these problems was in $\asOm n$.
So, it came as a surprise when Aaronson and Wigderson~\cite{AW08_Alg} demonstrated \MA-protocols for \Disj\ and \IP\ of cost $\asO{\sqrt n\log n}$, later improved by Chen~\cite{C18_On} to $\asO{\sqrt {n\log n\log\log n}}$.
That %
highlighted   %
the importance of proving the ``ultimate'' lower bound of $\asOm n$\ for the \MA-complexity of any explicit communication problem.

We will define a communication model \MAt\ (Def.~\ref{d_MAt}) that can be viewed as ``non-uniform \MA.''
Non-uniformity is the only possible source of advantage of \MAt\ over \MA:\ we will see (\clmref{c_MAunif}) that imposing the ``uniformity constraint'' on \MAt-protocols makes them not stronger than \MA-protocols.
All known lower bounds on $\MA(f)$ readily translate to $\MAt(f)$.
Intuitively, a lower bound argument that explores the uniformity of \MA\ (as opposed to \MAt) must have a very unusual structure.

We will see (\theoref{t_MAtAM}) that for any $f$ it holds that
$
\MAt(f) \in \asO{\sqrt{n\tm\AM(f)}}
;$
in other words, any lower bound of the form $\MAt(f) \in \asom{\sqrt n}$ will have non-trivial consequences for $\AM(f)$.\fn[fn_nlogn]
{
It is not too hard to demonstrate $\AM(f)\in\asOm{\log n}$ for an explicit $f$:\ for example, it holds for so-called index function $Ind(x,i)\deq x_i$; however, it is not clear how to use such examples to obtain $\MAt(f) \in \asOm{\sqrt{n\log n}}$.
}
Furthermore, according to \clmref{c_MAunif}, any lower bound of the form $\MA(f) \in \asom{\sqrt n}$ either should exploit the uniformity of \MA\ (the only difference between \MA\ and \MAt), or it will have non-trivial consequences for $\AM(f)$.
This partially explains why no such lower bound has been found yet.

\ndefi[d_MAt]{Non-uniform Merlin-Arthur, \MAt}{
For every $n\in\NN$, let $f:\OI^n\times\OI^n\to\set{0,1,\bot}$.

If for some $k(n)$, every input distribution $\mu_n$ and every $\eps>0$ there are functions $h_1\dc h_{2^{k(n)}}:\OI^n\times\OI^n\to\set{0,1,\bot}$, whose \Pp-complexity is in $\asO{k(n)\tm\log\fr1\eps}$, such that\mal{
\PR[(X,Y)\sim\mu_n]{f(X,Y) \neq \bigvee_{i=1}^{2^{k(n)}}h_i(X,Y)} \le \eps
,}
then we say that the \MAt-complexity of $f$, denoted by $\MAt(f)$, is in $\asO{k(n)}$.
}

The intuition behind the above formulation is the following.\fn
{
The author thanks the anonymous referee whose comment has resulted in the appearance of this paragraph.
}
Any \MA-protocol allows for error reduction at the cost of (at most) a multiplicative factor of $\asO{\log\fr1\eps}$, where $\eps$ is the ``target error.''
It has been intuitively clear that this this property of \MA\ is important:\ in particular, it was used by Klauck~\cite{K03_Rec} to prove his lower bound on the \MA-complexity.
The concept of \MAt-complexity \emph{isolates} this property, effectively allowing for its direct analysis, which is the main subject of this part of our work.

First of all, let us see that the non-uniformity is the only possible source of advantage of \MAt\ over \MA.

\clm[c_MAunif]{For every $n\in\NN$, let $g_1\dc g_{2^{k(n)}}:\OI^n\times\OI^n\to\set{0,1,\bot}$ be such that for every input distribution $\mu_n$ and every $\eps>0$ the conditions of \defiref{d_MAt} hold, as well as the additional requirement that
\mal{
\forall i\in\left[2^{k(n)}\right]:
~\PR[(X,Y)\sim\mu_n]{h_i(X,Y) \neq g_i(X,Y)} \le \eps
.}
Then the \MA-complexity of $f$ is in $\asO{k(n)}$.
}

Note that the functions $g_1\dc g_{2^{k(n)}}$ are fixed (for every $n$), in particular, they do not depend on $\mu_n$ $\eps$.
The statement says that in order to become sufficient for \MA, the definition of \MAt\ should be restricted by the additional requirement that all
the $h_i$  %
are approximations of the corresponding $g_i$.
That is why we view \MAt\ as a non-uniform modification of \MA.

\prfstart[\clmref{c_MAunif}]
Assume $\MAt(f)\in\asO{k(n)}$.
For every input distribution $\mu$ and $\eps>0$, let $h_i^{\mu,\eps}$ denote the function $h_i$ corresponding to these $\mu$ and $\eps$ from the definition of $\MAt(f)$.

Let $\nu$ be the uniform input distribution, then
\mal{
\forall x,y:~
\bigvee_{i=1}^{2^{k(n)}} h_i^{\nu,\eps}(x,y)
~\underset{\eps\to 0}{\longrightarrow}~
\bigvee_{i=1}^{2^{k(n)}} g_i(x,y)
}
and
\mal{
\forall x,y:~
\bigvee_{i=1}^{2^{k(n)}} h_i^{\nu,\eps}(x,y)
~\underset{\eps\to 0}{\longrightarrow}~
f(x,y)
,}
therefore
\mal[m_fgi]{
f(x,y)\equiv\bigvee_{i=1}^{2^{k(n)}}g_i(x,y)
.}

By the definition of \MAt\ it must hold that $\Pp(h_i^{\mu,\frac13})\in\asO{k(n)}$ for every input distribution $\mu$.
On the other hand,
\mal{
\forall \mu:~
\PR[(X,Y)\sim\mu]{h_i^{\mu,\fr13}(X,Y) \neq g_i(X,Y)} \le \fr13
,}
which means that $\BPP(g_i)\in\asO{k(n)}$.
Together with \bref{m_fgi} this implies $\MA(f)\in\asO{k(n)}$.
\prfend

Next we claim that a super-$\sqrt n$ lower bound on $\MAt(f)$ would have non-trivial consequences for $\AM(f)$.

\theo[t_MAtAM]{For any bipartite Boolean function $f$:
\mal{
\MAt(f) ~\in~ \asO{\sqrt{n\tm\AM(f)}}
.}
}

\prfstart[\theoref{t_MAtAM}]
Let $\AM(f)=k(n)$ -- that is, for every input distribution $\nu$ there is an \NP-protocol of cost at most $k(n)$ that solves $f$ with error at most $\frac13$ with respect to $\nu$.
Via the standard accuracy amplification technique this implies that for any input distribution $\nu$ and $\eps>0$ there is an \NP-protocol of cost $\asO{k(n)\tm\log\frac1\eps}$ that solves $f$ with error at most $\eps$ with respect to $\nu$.
In particular, for every input distribution $\nu$ there is an \NP-protocol $\Pi_\nu$ of cost $\asO{\sqrt{n\tm k(n)}}$ that solves $f$ with error at most $2^{-\sqrt{\frac n{k(n)}}}$ with respect to $\nu$.

Let us see that
\mal[m_wishful]{
\MAt(f)\in\asO{\sqrt{n\tm k(n)}}
.}
For $n\in\NN$, take any input distribution $\mu_n$ and any $\eps>0$.

If $\eps\le2^{-\sqrt{\frac n{k(n)}}}$, then let $h_1=f$:\ as its \Pp-complexity is at most $n\in\asO{\sqrt{n\tm k(n)}\tm\log\frac1\eps}$, the ``decomposition''
\mal{
f(X,Y) = \bigvee_{i=1}^{1}h_i(X,Y)
}
satisfies the requirements of \defiref{d_MAt} with respect to \bref{m_wishful}.

Now suppose that $\eps>2^{-\sqrt{\frac n{k(n)}}}$.
Then $\Pi_\mu$ is an \NP-protocol of cost $\asO{\sqrt{n\tm k(n)}}$ that solves $f$ with error less than $\eps$ with respect to $\mu$.
Let $K\in2^{\asO{\sqrt{n\tm k(n)}}}$ be the number of rectangles contained in $\Pi_\mu$, denote their characteristic functions by $h_1\dc h_K$.
As the \Pp-complexity of every such $h_i$ is $1$ and
\mal{
\PR[(X,Y)\sim\mu_n]{f(X,Y) \neq \bigvee_{i=1}^Kh_i(X,Y)}
= \PR[\mu_n]{f(X,Y) \ne \Pi_\mu(X,Y)}
< \eps
,}
the requirements of \defiref{d_MAt} with respect to \bref{m_wishful} are satisfied, and the result follows.
\prfend

\sect[s_conc]{Conclusions}

Among those communication complexity regimes that reside well beyond our current level of understanding, the model of \emph{Arthur-Merlin (\AM)} may be the closest to us.
The motivation of this work has been to explore the ``neighbourhood'' of \AM\ that we might be able to analyse.

\itemi{
\item We have defined and analysed a new communication complexity
class,  %
\SLAM, strictly included in \AM\ and strictly stronger than the union
of all previously known %
subclasses of \AM.  %
\item We have identified one possible source of hardness in proving $\asom{\sqrt n}$ lower bounds against \MA:\ such a bound would either be of a ``very special form,'' or imply a non-trivial lower bound against \AM.
}

A few questions that have remained open can be viewed as possible further steps towards understanding \AM.
For instance:
\itemi{
\item What is the \SLAM-complexity of \Disj?
Note that even its \UAM-complexity is not understood yet (see~\cite{GPW15_Zero} for details).
\item Can we prove a lower bound of $\asOm{\sqrt{n\log n}}$ on the \MA-complexity of an explicit function (see Footnote~\ref{fn_nlogn})?
\item What approaches to understanding \AM\ look promising?
\itemi{
\item Shall we try hard to prove a lower bound of $n^{\frac12+\asOm1}$ on the \MA-complexity of an explicit function?
\item Are there %
complexity classes inside \AM\
with non-trivial advantage over \SLAM\ (or incomparable to it), which we can analyse?%
}
}

Finally, we would like to mention a result that is somewhat dual to this work from the conceptual point of view:\ Bouland, Chen, Holden, Thaler and Vasudevan~\cite{BCHTV17_On} define communication complexity classes $\mathit{NISZK^{cc}}$ and $\mathit{SZK^{cc}}$, and show that
\mal{
\mathit{NISZK^{cc}} \sbseq \mathit{SZK^{cc}} \sbseq \AM
\txt{~~~and~~~}
\mathit{NISZK^{cc}} \nsbseq \UPP
,}
where \UPP\ is the class consisting of functions with \emph{high sign-rank}; \UPP\ is known to strictly contain \PP.
Accordingly, the quest of understanding \AM\ is at least as hard as that of understanding $\mathit{NISZK^{cc}}$, and the latter might be simpler if $\mathit{NISZK^{cc}}\sbs\AM$. 

\sect*{Acknowledgements}

I am grateful to Thomas Watson for many useful comments and suggestions.
I would also like to thank the authors of helpful anonymous reviews.
The most careful proofreading done by the editorial team of \e{Theory of Computing} has contributed not just to the presentation quality of the results, but also to the author's confidence in them.

\newcommand{\etalchar}[1]{$^{#1}$}

\end{document}